# ASSESSMENT OF A FAILURE PREDICTION MODEL IN THE ENERGY SECTOR: A MULTICRITERIA DISCRIMINATION APPROACH WITH PROMETHEE BASED CLASSIFICATION

Silvia Angilella[1], Maria Rosaria Pappalardo[2]

Department of Economics and Business, University of Catania, Corso Italia, 55, Catania 95129, Italy.
1 Email: silvia.angilella@unict.it
2 Corresponding author. Email: mrosaria.pappalardo@unict.it

**Abstract**

This study presents the implementation of a non-parametric multiple criteria decision aiding (MCDA) model, the Multi-group Hierarchy Discrimination (M.H.DIS) model, with the Preference Ranking Organization Method for Enrichment Evaluations (PROMETHEE), on a dataset of 114 European unlisted companies operating in the energy sector.
Firstly, the M.H.DIS model has been developed following a five-fold cross validation procedure to analyze whether the model explains and replicates a two-group pre-defined classification of companies in the considered sample, provided by Bureau van Dijk's Amadeus database. Since the M.H.DIS method achieves a quite limited satisfactory accuracy in predicting the considered Amadeus classification in the holdout sample, the PROMETHEE method has been performed then to provide a benchmark sorting procedure useful for comparison purposes.
The analysis indicates that in terms of average accuracy, M.H.DIS model development with the PROMETHEE based classification provides consistently better results compared to the one obtained with the Amadeus classification in the majority of combinations, which have been built with the financial variables covering the main firm's dimensions such as profitability, financial structure, liquidity and turnover.

**Keywords:** Multiple Criteria Decision Aid, Credit rating, Energy companies, Multi-group Hierarchical Discrimination (M.H.DIS) model, Preference Ranking Organization Method for Enrichment Evaluations (PROMETHEE).

## 1. INTRODUCTION

Tracking economy-wide energy management is a crucial issue as confirmed by several actions, undertaken by policymakers in many countries, alongside different dimensions: financial, environmental, sustainability, technical and market conditions (Angilella & Pappalardo, 2020). Therefore, energy companies play a key role in the whole economy of a country, providing essential services to all sectors and enhancing the living standards and the socio-economic growth of a country. In last decades, some energy companies faced severe financial soundness issues due to flawed risk management actions of banks and deregulation processes introduced in the European energy industry on December 1996. For instance, the recent directives implemented to liberalize the European electricity sector, were aimed to lower consumers' prices and to create a more competitive context (Kočenda & Čábelka, 1998; Meyer, 2003). However, for the specific features of the energy sector such as the large infrastructure costs and the economic difficulties to replicate the transmission lines, the wholesale prices of energy market remained significantly high and their associated sales volume low. Hence, several financial distresses have occurred within the energy sector.

Various case studies across the world give examples of energy companies being challenged by deregulation processes, such as the British Energy company in 2000, the Electricaribe company of northern Colombia during the period from 2015 to 2016 and the California distribution companies from 2000 to 2001. These failures generated serious effects on the economy of a country, which have been promptly faced by governments' interventions, such as large expenses on the whole country's economy where the crash has taken place.

In order to prevent domino effects on the economy, it is important to monitor energy companies in terms of their financial performances.

In this regard, after deregulation processes, several studies have been mainly focused on the issue of assessing market (Denton et al., 2003), financial (Bjorgan, 1999) and price risk (Dahlgren et al., 2003). Despite the relevance of the topic, few studies have been specifically devoted to credit risk assessment of companies operating in this sector; the only one is the work of Silva and Pereira (2014) that assesses the credit risk of thirty renewable energy companies sited in Portugal employing a traditional linear regression model.

In this sector more than in others, the need for reliable credit risk assessment models able to predict corporate failure consistently and accurately, is crucial. In literature, methods mainly dealing with corporate failure prediction problems include statistical, econometric and machine learning techniques. In this regard, the literature review of Balcaen and Ooghe (2006) provides a well-organized overview of the classical statistical modelling systems applied to business failure predictions of corporations throughout 35 years of studies. This paper identifies specifically four types of approaches with their main features and assumptions: univariate analysis, risk index models, multiple discriminant analysis (MDA) and conditional probability models.

Despite their extensive implementation, these methodologies present some specific issues related to the application of corporate failure prediction modelling and do not hold some significant aspects that analysts often require scoring models to incorporate, such as the ordinal risk grades and the monotonicity assumptions. The last requirement entails that if in a rating model an input variable for a given firm improves, then the probability of default should decrease. All these attributes fit instead well to multi-criteria decision aiding (MCDA) models, which have also the advantages of a high comprehensibility, easiness of application and ability to include the DM's preferences. These characteristics make these tools more efficient and powerful than traditional statistical techniques (Doumpos et al., 2002).

MCDA are well-established methodologies for supporting decision-making problems in evaluating a set of alternatives under multiple and conflictual criteria (Greco et al., 2016).

Very often, MCDA models have been adopted to support financial decisions such as the portfolio selection, the choice of investments projects, the failure risk assessment of corporations (see Spronk et al., 2005 and Doumpos & Zopounidis, 2014 for literature reviews of MCDA on finance). Specifically, the last topic falls within the framework of sorting models employed to classify alternatives in two or more groups already defined.

In this regard, the purpose of this research is to develop a Multi-criteria decision aiding (MCDA) model for credit risk analysis of a set of European unlisted companies operating in the energy sector. However, different issues arise to deal with this aim: (a) which is the best performing MCDA model in terms of overall accuracy rate among various multi-criteria methods adopted in financial decision-making problems to predict the companies' failure risk? (b) which is the MCDA model fitting well to the financial evaluation of energy companies' performance? (c) Are these models robust? (d) how to support the credit risk assessment process for unlisted energy companies not rated by Credit Rating agencies (CRAs)?

Bearing in mind the above questions, the following considerations provide useful explanations.

MCDA offers a variety of discrimination models (see Zopounidis & Doumpos, 2002a for a literature review of multicriteria classification and sorting methods), which have been applied to handle with the credit risk assessment issue especially in financial and banking sector. Most of them make use of value functions (Zopounidis & Doumpos, 1999; Doumpos & Pasiouras, 2005; Baourakis et al., 2009), goal programming (Garcia et al., 2013), rough set theory (Slowinski & Zopounidis, 1995; Capotorti & Barbanera, 2012) and outranking techniques (Doumpos & Zopounidis, 2011; Angilella & Mazzù, 2015; Angilella & Mazzù, 2019).

In several credit risk assessment studies, some of these methods have been compared to each other (Araz & Ozkarahan, 2005) and with traditional econometric tools such as discriminant, logit and probit analysis (Voulgaris et al., 2000; Zopounidis & Doumpos, 1999). All these studies agree in recognizing the higher efficiency of multi-criteria methods in comparison to the econometric ones in obtaining credit risk estimates (Doumpos & Zopounidis, 2002). Instead, a more controversial question is about which multi-criteria model is more efficient in corporate credit risk assessment, because of the significant link between the features of the context of application and the obtained results.

In literature, one of the most efficient multi-criteria discrimination model is the Multi-group Hierarchy Discrimination (M.H.DIS) technique elaborated by Zopounidis and Doumpos (2000). In comparison to other studies concerning the application of preference disaggregation approaches, (such as the family of UTADIS models), the performance of M.H.DIS is indeed not only superior for some real world cases, but also computationally less time-consuming, especially with respect to UTADIS II and UTADIS III (1 minute against several hours) (Zopounidis and Doumpos, 2000).

The following features emphasize the M.H.DIS model's main strengths:

- it is able to discriminate alternatives between two or more than two categories;
- it employs a hierarchical discrimination procedure to assign alternatives into classes. More specifically, the categories are discriminated progressively, starting by discriminating the most preferred alternatives ($C_1$) from all the alternatives of the remaining ones $(C_2, C_3, C_4, \ldots, C_p)$ then proceeding to the discrimination between the alternatives of the next category ($C_2$) from all the alternatives of the remaining ones $(C_3, C_4, \ldots, C_p)$ and so forth;
- his development process is based on three mathematical programming techniques, two linear programs (LP1, LP2) and a mixed-integer one (MIP), implemented at each stage of the hierarchical discrimination process to estimate the optimal pair of additive utility functions in terms of misclassification errors and clear distinction between categories.

M.H.DIS model has been applied to several fields such as: the banking system (Pasiouras et al., 2010; Spathis et al., 2004), the corporate sector (Doumpos et al., 2002; Kosmidou et al., 2002; Doumpos & Zopounidis, 1999) and the country analysis (Doumpos & Zopounidis, 2001; Doumpos et al., 2000). However, to the best of our knowledge, the M.H.DIS model has never been applied to financial distress prediction of energy companies despite their great impact on country's economy.

Furthermore, different multicriteria models have been implemented in the energy sector (see for literature reviews on this topic Pohekar & Ramachandran, 2004; Mardani et al., 2015). However, most of them have been related on some specific aspects regarding: the choice of the power plants location (Choudhary & Shankar, 2012; Wu et al., 2014), and evaluation (Atmaca & Basar, 2012; Liu et al., 2010); the project selection (Chen et al., 2010; Lee et al., 2009), the choice of the future energy supplier (Fittipaldi et al., 2001), the sustainability assessment of electricity production (Troldborg et al., 2014) and supply technologies (Hirschberg et al., 2004).

The branch of MCDA literature closely related to credit risk assessment is the one on firms' performance evaluation (Psillaki et al., 2010). This last topic has been only dealt with recent papers by Eyüboglu & Çelik (2016) and Angilella & Pappalardo (2020) considering different companies' dimensions such as: financial, technical, environmental and market condition.

Thus, in order to fill the gap in multicriteria sorting models employed to evaluate the credit risk of energy companies, the aim of this research is to apply the Multi-group Hierarchical Discrimination (M.H.DIS) model of Zopounidis and Doumpos (2000) to a balanced sample of 114 active and inactive energy companies for up to four years prior the financial distress occurred. In order to observe whether a pre-defined classification of companies in two categories, active and inactive ones, provided by Bureau van Dijk's Amadeus database is well replicated by the model, a five-fold cross validation has been performed on companies of the sample.

Despite what we expect, the average accuracy rate of the M.H.DIS model developed on Amadeus classification is not quite satisfactory in the holdout sample of the analysis. Therefore, in this study we consider a further well-known multicriteria decision aid model, the Preference Ranking Organization Method for Enrichment Evaluations (PROMETHEE II) (presented for the first time in Mareschal et al., 1984), on which a classification of firms in the sample is based. Such classification acts as benchmark sorting on which to compare the accuracy of the discrimination model.

To deal with this aim, we identify first six financial ratios that in our analysis result the most powerful in highly discriminating between the two categories of companies. Then, they have been considered in all the possible combinations of subsets of three criteria and employed in turn in the building of PROMETHEE II classification first and M.H.DIS model development then.

Thus, the contribution of this paper is fourfold:

- we address the literature gap in multicriteria sorting models, enriching applications of the M.H.DIS model also on credit risk assessment of energy companies;
- we extend the M.H.DIS model development with a well-established multicriteria outranking model, the PROMETHEE II method, to provide a benchmark sorting procedure to compare with the pre-defined classification given by Amadeus database;
- we suggest a novel use of the proposed discrimination model to support the credit risk assessment process for firms lacking of a synthetic judgment provided by credit rating agencies (CRAs);
- we provide a more consistent and robust discrimination model in terms of average and overall accuracy rate. The robustness is examined over time, under different combinations of financial variables, under different preference functions employed for PROMETHEE classification and simulating the criteria weights in different scenarios.

The rest of the paper is organized as follows. Section 2 introduces all the basic notions relative to M.H.DIS model and PROMETHEE II method. Section 3 presents the data used in the analysis and the sampling procedure. Section 4 discusses about the building of M.H.DIS model starting from a three steps selection procedure of the most predictive variables in discriminating between active and inactive companies. Section 5 shows the results of the M.H.DIS model in terms of accuracy for the AMADEUS classification, whereas Section 6 develops a specific classification of companies according the PROMETHEE II method for comparison purposes. Section 7 summarizes the main findings. Section 8 concludes the paper and discusses some future research directions.

## 2. BASIC CONCEPTS

In this section, we give some basic concepts used further in the paper. In Section 2.1 we present an overview the Multi-group Hierarchical Discrimination (M.H.DIS), while in Section 2.2 we recall the Preference Ranking Organization Method for Enrichment Evaluations (PROMETHEE).

2.1 THE MULTI-GROUP HIERARCHICAL DISCRIMINATION MODEL (M.H.DIS)

MCDA provides a variety of different models that help a decision maker (DM) to solve three main problems: choice, ranking or sorting. Choice problem consists of the selection of a subset of alternatives from a given initial set of options; ranking problem requires to rank alternatives in a partial or total order, while in sorting problems each alternative has to be assigned to one or more contiguous preferentially ordered classes (for a comprehensive taxonomy of a MCDA process see Cinelli et al., 2020).

Generally, in a MCDA problem (Figueira et al., 2005), there is a finite set of alternatives $A = \{a_1, \cdots, a_j, \cdots, a_m\}$, which are evaluated on a consistent family of criteria $G = \{g_1, \cdots, g_i, \cdots, g_n\}$.

In this study, we employ the M.H.DIS model developed by Doumpos and Zopounidis (2000), to solve the sorting problem of the assignment of a given set of alternatives into predefined ordered classes. More in detail, we will use the following notation:

- $A = \{a_1, \cdots a_j, \cdots, a_m\}$ is the set of finite alternatives;
- $G = \{g_1, \cdots g_i, \cdots, g_n\}$ is the set of consistent criteria with an increasing or decreasing direction of preference order;
- $a_{ji}$ indicates the evaluation of alternative $j$ on criterion $i$;
- $C = \{C_1 \succ \cdots \succ C_k \cdots \succ C_p\}$ is the set of $p$ ordered categories from the best (or healthiest) $C_1$ to the worst (or riskiest) $C_p$.

In this analysis, the set of alternatives $A$ is composed of 114 European unlisted companies operating in the energy sector. The aim of M.H.DIS model is to sort these companies into two predefined categories, the active and the inactive ones, respectively denoted with $C_1$ and $C_2$.

Alternatives are evaluated on a set of criteria $G$ representing the main financial aspects of companies and endowed of a high predictor power in distinguishing between active and inactive companies. Moreover, for simplicity of computation, the model has been implemented only in the case in which criteria present an increasing preference direction, implying that the evaluation of a company on an attribute $g_i$ that is negatively (positively) related to financial distress, increases its likelihood to be assigned to the best (worst) category.

Furthermore, M.H.DIS model is a credit risk assessment technique, such as discriminant, logit and probit analysis, that requires two distinct samples to be applied: a basic sample (training set) to build a model able to reproduce the pre-specified classification as much as possible, and a holdout sample (test set) to validate and verify its generalization of application. Hence, also the following two subsets of $A$ have to be considered in the building of M.H.DIS model:

- $B = \{b_1, \cdots b_r, \cdots, b_s\}$ is the subset of alternatives composing the training sample, used for model development;
- $D = \{d_1, \cdots d_s, \cdots, d_t\}$ is the subset of alternatives composing the test sample, used for validation purposes with $B \cap D = \emptyset$.

Initially, the alternatives of the training sample are evaluated on the attributes in $G$ and each of them is assigned to a pre-specified category $C_k$; once it is carried out, the model aims to sort companies into two categories in order to replicate, as much as possible, a given classification before model development. Then, the discriminating procedure is applied also to companies of test sample to classify them and validate the results.

In order to sort companies of training set, M.H.DIS model applies the following hierarchical technique. The procedure starts from stage $k = 1$ by considering the best category $C_1$ to which companies of training set ($b_r$) can belong. In $k = 1$, the model builds a pair of additive utility functions, of which formulas are provided below, to discriminate companies belonging to the healthiest category $C_1$ and companies belonging to the remaining riskier categories than $C_1$ (i.e. $C_2$ in our context):

$$U_1(\bar{g}(b_r)) = \sum_{i=1}^{n} h_1 u_{1i}(g_i(b_r)), \qquad (1)$$

$$U_{\sim 1}(\bar{g}(b_r)) = \sum_{i=1}^{n} h_{\sim 1} u_{\sim 1i}(g_i(b_r)), \qquad (2)$$

where $U_1(\bar{g}(b_r)) \in [0,1]$ and $U_{\sim 1}(\bar{g}(b_r)) \in [0,1]$ represent the two additive utility functions of each alternative $b_r$; $\bar{g}$ is the global evaluation of each alternative ($b_r$) on the whole set of criteria considered; $u_{1i}(g_i(b_r))$ and $u_{\sim 1i}(g_i(b_r))$ indicate the estimated two marginal utility functions with an increasing (or decreasing) preference direction according to each attribute $g_i$ negatively (or positively) related to financial distress; $h_1$ and $h_{\sim 1}$ denote the weights of each criterion summing to one.

In stage $k = 1$, if the global score of the estimated additive utility function of healthiest category for alternative $b_r$, is higher than the global score of the estimated additive utility function of the riskiest categories, i.e. $U_1(\bar{g}(b_r)) \geq U_{\sim 1}(\bar{g}(b_r))$, then $b_r$ is classified to category $C_1$; otherwise if $U_1(\bar{g}(b_r)) \leq U_{\sim 1}(\bar{g}(b_r))$, company $b_r$ does not belong to class $C_1$ and the procedure will continue to stage $k = 2$. From stage 1, it has to be highlighted that if the strict inequality among the global scores of the estimated utility functions occurs $\left(U_1(\bar{g}(b_r)) > U_{\sim 1}(\bar{g}(b_r))\right)$, then company $b_r$ is classified correctly by the model; on the contrary if the two estimated additive utility functions are equal $\left(U_1(\bar{g}(b_r)) = U_{\sim 1}(\bar{g}(b_r))\right)$, then the model misclassifies the company. The whole set of companies correctly or incorrectly classified in $C_1$ by the model, are excluded in next stages.

At stage $k = 2$, analogously the model builds another pair of additive utility functions to discriminate companies belonging to category $C_2$ from companies belonging to the remaining riskier categories than $C_2$ (i.e. $C_3, C_4, \cdots, C_p$). Similarly to stage 1, if $U_2(\bar{g}(b_r)) \geq U_{\sim 2}(\bar{g}(b_r))$ or $U_2(\bar{g}(b_r)) \leq U_{\sim 2}(\bar{g}(b_r))$, then company $b_r$ is classified respectively into $C_2$ or $C_{\sim 2}$.

The same discriminating procedure continues until all companies of training sample have been classified into the ordered categories to replicate the pre-specified classification as much as possible. M.H.DIS model is also applied to companies of test sample in the same manner.

Figure 1 shows the hierarchical discrimination technique employed to perform the M.H.DIS model. In order to generalize the hierarchical discriminating procedure to $p$ categories, the expressions (1) and (2) are replaced with the following:

$$U_k(\bar{g}(b_r)) = \sum_{i=1}^{n} h_k u_{ki}(\bar{g}(b_r)), \qquad (3)$$

$$U_{\sim k}(\bar{g}(b_r)) = \sum_{i=1}^{n} h_{\sim k} u_{\sim ki}(\bar{g}(b_r)) \qquad (4)$$

Hence, the model will build as many pairs of additive utility functions as $p - 1$ classes to which companies have to be sorted.

Furthermore, to estimate optimally the additive utility functions of the model at each stage k, two mathematical programing techniques have been solved through a Matlab code: two linear programs (LP1 and LP2) and a mixed-integer program (MIP). The linear program LP1 has been implemented with the mixed-integer program MIP first, to minimize the misclassification costs of companies belonging to other categories than the pre-defined one; the second linear program LP2 has been performed then, to enhance the clarity of the obtained classification as an among-group variance maximization in discrimination analysis. Further details on the assessment of the additive utility functions can be found in Zopounidis and Doumpos (2002b).

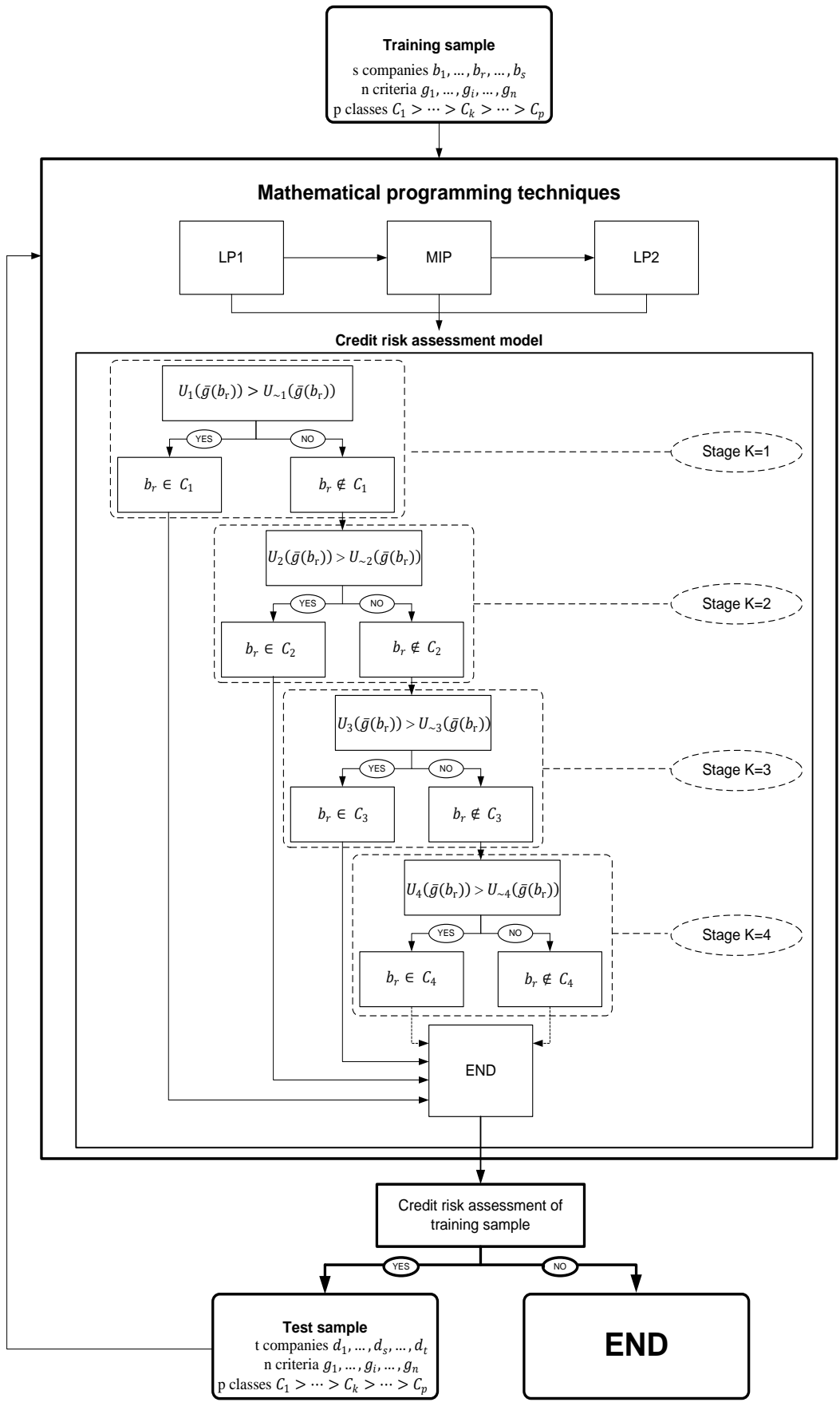

Figure 1. General scheme of model development in the M.H.DIS model. Authors' elaboration

## 2.2 PROMETHEE II

PROMETHEE II is a multi-criteria method belonging to the family of PROMETHEE methods, that builds an overall composite indicator of alternatives on the basis of pairwise comparisons considering a set of criteria. PROMETHEE methods are widely used in Multiple Criteria Decision Aiding (see Brans & De Smet, 2016 for a state-of-the-art on the topic). There is a considerable number of PROMETHEE applications currently available for various fields. With respect to financial topics, PROMETHEE methods have been already successfully applied for example in banking (Mareschal & Brans, 1991 and Doumpos & Zopounidis, 2010), in asset evaluation (Albadvi et al., 2007), in bankruptcy prediction (Hu & Chen, 2011 and Mousavi & Lin, 2020), in portfolio selection (Vetschera & de Almeida, 2012), in country risk assessment (Doumpos & Zopounidis, 2001) and in performance assessment of microfinance institutions (Gaganis, 2016).

Among the different versions of PROMETHEE methods, PROMETHEE II is the most frequently applied one because it enables a decision maker (DM) to obtain a complete ranking of the alternatives. It is based on the preference function $P_i(a_j, a_h)$ representing the degree of preference of alternative $a_j$ on $a_h$. $P_i(a_j, a_h)$ is a non-decreasing function of the difference $d_i = g_i(a_j) - g_i(a_h)$. In Mareschal, Brans and Vincke (1984), the multi-criteria methodology PROMETHEE II has been presented, considering six different types of preference functions: the regular criterion, the u-shape-criterion, the v-shape criterion, the level criterion, the criterion with linear preference and indifference area and the Gaussian criterion.

In this study, each preference function is employed to build a binary classification of companies which will be compared with the one provided by AMADEUS database. Moreover, the whole set of preference functions is used to observe how the classification made with PROMETHEE II method varies according the type of function considered (Figure 2).

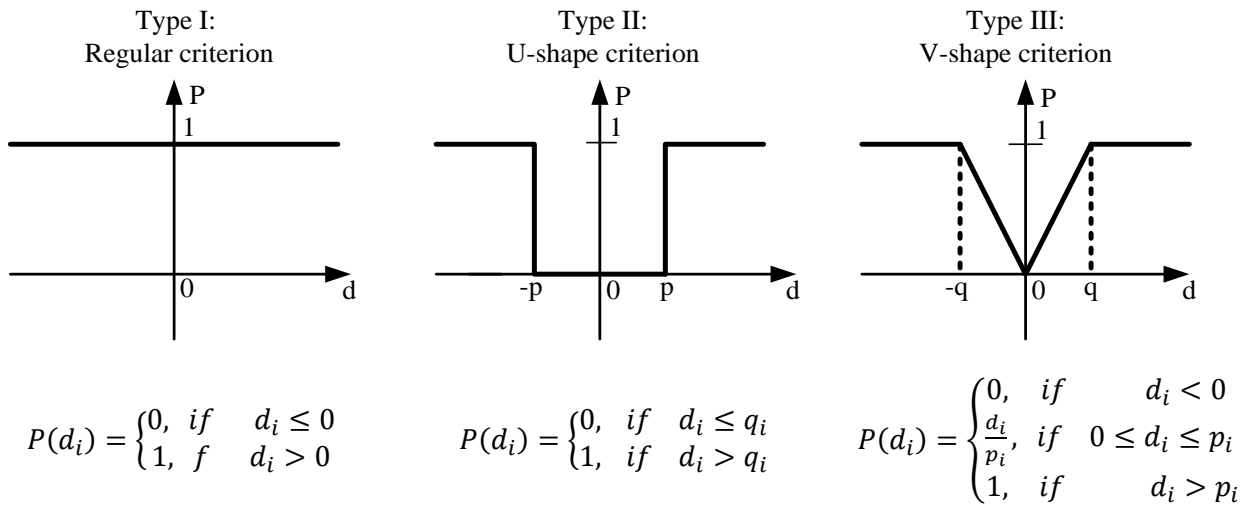

Type I: Regular criterion

$$P(d_i) = \begin{cases} 0, & if \quad d_i \leq 0 \\ 1, & f \quad d_i > 0 \end{cases}$$

Type II: U-shape criterion

$$P(d_i) = \begin{cases} 0, & if \quad d_i \leq q_i \\ 1, & if \quad d_i > q_i \end{cases}$$

Type III: V-shape criterion

$$P(d_i) = \begin{cases} 0, & if \quad d_i < 0 \\ \frac{d_i}{p_i}, & if \quad 0 \leq d_i \leq p_i \\ 1, & if \quad d_i > p_i \end{cases}$$

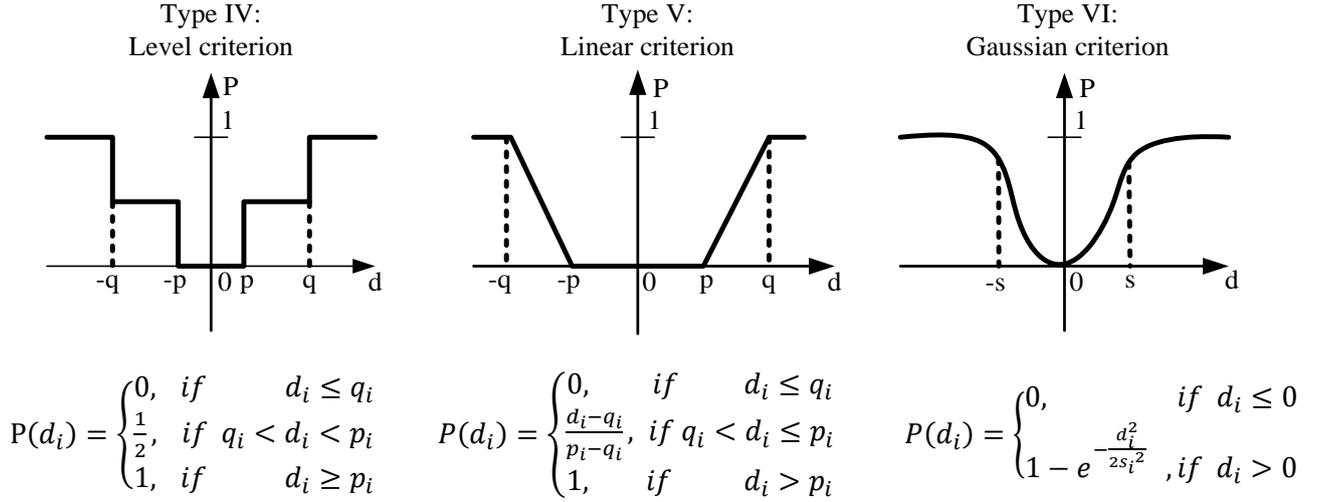

Figure 2. Types of preference functions $P(d_i)$. Authors' elaboration

Considering for each criterion $g_i$ a weight $w_i$ such that $w_i \geq 0$ and $\sum_{i=1}^{n} w_i = 1$, PROMETHEE II computes:

$$\pi(a_j, a_y) = \sum_{i=1}^{n} w_i P_i(a_j, a_y), \qquad (5)$$

which represents the strength of preference of alternative $a_j$ with respect to $a_y$ on the basis of the whole set of criteria. Then PROMETHEE II compares each alternative $a_j$ with the other alternatives by computing the positive and negative inflow of $a_j$, defined, respectively, as follows:

$$\Phi^+(a_j) = \frac{1}{m-1} \sum_{a_j \in A \setminus \{a_y\}} \pi(a_j, a_y), \text{ and } \Phi^-(a_j) = \frac{1}{m-1} \sum_{j \in A \setminus \{a_y\}} \pi(a_y, a_j). \qquad (6)$$

Finally, PROMETHEE II builds a net flow for each alternative by:

$$\Phi(a_j) = \Phi^+(a_j) - \Phi^-(a_j). \qquad (7)$$

PROMETHEE II provides a complete preorder on $A$ ranking the alternatives from the best to the worst. The net flow takes values in the range $[-1,1]$; if $\Phi(a_j) \simeq 1$, then $a_i$ is almost strictly preferred over all alternative, while if $\Phi(a_j) \simeq -1$, then $a_j$ is almost strictly preferred by all the alternatives. In this paper the following assumptions have been considered:

- the weights of criteria have been simulated using a hit and run procedure (see Smith, 1984) with 10,000 scenarios similarly to SMAA-PROMETHEE method (Corrente et al., 2014) but without providing preference information of Decision Maker on the parameters involved and without estimating the SMAA indices; under each scenario the net flow of each company has been evaluated and the average net flow of each company has been computed for all the scenarios;
- the user constant $s_i$ of the Gaussian criterion, has to be determined on the basis of a rule of thumb: $s_i = \frac{p_i + q_i}{2} > 0$, where for each $g_i \in G$, $p_i$ and $q_i$ are, respectively, the preference and indifference thresholds;

- as in Rogers and Bruen (1998), for each $g_i \in G$ we have assumed $p_i$ and $q_i$ constant, computed as follows: $p_i = \frac{2}{3} r_i$ and $q_i = \frac{1}{6} r_i$ with $r_i = |\max(g_i) - \min(g_i)|$.

## 3. DATA COLLECTION

The first step to develop a risk assessment model is the selection of the firms' sample. The primary source for the collection was the database of Bureau van Dijk's Amadeus. We have looked for unlisted companies with the NACE code 35 used as filter, which covers the main industrial sectors in the energy sector. Within the NACE code 35, we have chosen specifically the code 351, indicating electricity, gas, steam and air conditioning supply sector, articulated in the electricity production, transmission, distribution and trade segments. Among the unlisted companies operating in the energy supply chain, only those located in the 28 countries of the European Union, with active and inactive status, have been selected. An inactive company is defined by Amadeus like the one that is in liquidation, bankruptcy or dissolved (merger, take-over, demerger). Thus, the original sample consists of 219 inactive companies and 5736 active companies that has been further cleaned up of all missing values also in relation to the identified events of financial distress spanning the period 2013-2018 chosen in this study. Hence, inactive and active companies have been reduced respectively to 57 and 1551. Then a stratified resampling method, consisting of deriving the same number of failed and non-failed firms by matching them to the inactive ones of the same size, has been applied to the original sample, to avoid problems of inconsistent parameter estimation and under-valuate misclassification error rate that may arise with an unbalanced sample. According to a careful screening process, the 1551 active companies have been classified into four groups (large, medium, small and micro-companies) with respect to their size, which is traditionally measured by three parameters: number of employees, annual revenues and annual assets. The four groups have been labelled as follows: Large (1), Medium (2), Small (3) and Micro (4). Thus, the number of active companies, denoted as $Active_j$ with $j \in \{1,2,3,4\}$, and belonging to each category, has been reduced by applying the following formula:

$$Active_j = Tot.\,Inactive \cdot \left(\frac{n_j}{Tot.\,Active}\right), \qquad (8)$$

with $n_j$ denoting the number of large, medium, small and micro active companies, $Tot.\,Inactive$ and $Tot.\,Active$ indicating, respectively, the total number of inactive and active companies.

Thus, the final sample constructed through the above procedure involves 28 countries and consists of 114 unlisted European energy companies.

Table 1 shows the balanced sample classified into inactive and active companies obtained after the stratified resampling method of 1551 active companies, whereas Table 1A in the Appendix displays the set of 114 Energy companies distributed per country.

Table 1. Balanced sample after the stratified resampling method. Authors' elaboration

| STATUS OF ENERGY COMPANIES | SIZE OF COMPANIES | | | | TOTAL |
| --- | --- | --- | --- | --- | --- |
| | LARGE | MEDIUM | SMALL | MICRO | |
| INACTIVE | 28 | 8 | 17 | 4 | **57** |
| ACTIVE | 827 | 635 | 83 | 6 | **1551** |
| ACTIVE AFTER RESAMPLING METHOD | 30.392 | 23.336 | 3.050 | 0.220 | 57 |
| ACTIVE$_j$ | 30 | 23 | 3 | 1 | **57** |

Moreover, a five-fold cross-validation has been performed in order to eliminate the problem of small sample and to develop the model adequately. Thus, the final balanced sample consisting of 114 energy companies has been split, in a random way, into five mutually exclusive folds of equal size composed respectively of training and test set in the proportion of 80% and 20%. Each fold contains a training set of 92 companies to fit the model and a test set of 22 companies for validation purposes. The average accuracy rate over all the five folds is the cross-validated accuracy rate.

## 4. M.H.DIS MODEL BUILDING

Once the final sample has been balanced, the development of credit risk assessment model requires a careful selection of predictor variables able to well discriminate among active and inactive companies. In the next section, a literature review of independent variables most widely employed in failure prediction models is discussed, whereas Section 4.2 involves a careful screening in three steps, able to detect variables with a high explanatory and discriminating power between active and inactive companies.

### 4.1 INDEPENDENT VARIABLES SELECTION: LITERATURE REVIEW IN FAILURE PREDICTION MODELS.

A large growing body of literature on failure prediction models based on Financial Ratios (FRs) is available for their easiness of assessment from the financial statements (Altman, 1968; Beaver, 1966; Ohlson, 1980). Usually, the employed FRs have been grouped according to the main firms' dimensions such as profitability, financial structure, liquidity, solvency, turnover and activity, which provide insights on how companies' internal aspects affect their risk of failure.

Generally, scholars have adopted a wide range of predictor variables in numerous scientific research studies; however, analytical predictive models have to comply with a tradeoff: a limited set of predictors to fit the model, able to represent all relevant information without creating overlapping, together with a low over-fitting on the training sample and a high performance on the test sample.

For these reasons, a careful screening process has to be performed to provide more accuracy in the distress prediction model.

This section deals with one of the initial steps in the development of a failure prediction model: the selection of the most predictive variables. It consists of the review of the distress prediction literature with special attention to the most predictive variables employed in the energy sector (see for a literature review of FRs on failure prediction models: Xu et al., 2019; Liang et al., 2016; Du Jardin, 2016).

Table 2 shows the set of 42 FRs derived from the literature review, which are classified according to six firms' dimensions, together with their acronyms and definitions. The list includes also other FRs measuring the company size, because some scholars suggest that dimensional difference among companies is a key factor affecting the company's default probability. To this aim, we include two other variables widely used in literature, namely total assets and total sales revenue as proxies for the firm size (Al-Khazali & Zoubi, 2005). Variables denoted with * have been eliminated from Table 2 because of abridgement of the information provided by the financial statements of Amadeus. Then the total number of FRs used for subsequent analysis has been reduced to 37.

Table 2. Financial ratios derived from literature review of failure prediction models.

| FINANCIAL RATIOS (FRs) | | | |
|---|---|---|---|
| Acronym | Variables | Definition | Tot. |
| **PROFITABILITY** | | | |
| EBIT_TA | EBIT/tot. assets | Ebit/tot. assets | |
| LTDR | Long-term debt ratio | Long-term debt/tot. assets | |
| OP_MARG | Operating margin | EBIT/net sales | |
| PROF_MARG | Profit margin | Net income/net sales | |
| ROE | ROE | Net income/stareholders' equity | |
| ROA | ROA | Net income/ tot. assets | 11 |
| ROCE | ROCE | EBIT/(currents assets-current liabilities) | |
| EBIT_EQ | EBIT/shareholder funds | EBIT/shareholder funds | |
| EBITDA_TA | EBITDA/tot. assets | EBITDA/tot. assets | |
| CF_TA | Cash flow/tot. assets | Cash flow/tot. assets | |
| CF_EQ | Cash flow to equity | Cash flow/shareholders' equity | |
| **FINANCIAL STRUCTURE** | | | |
| EQ_RATIO | Equity ratio | Tot. equity/tot. assets | |
| FAT | Fixed asset turnover | Net sales/ fixed assets | |
| *IC | *Interest coverage | *EBIT/interest expense | |
| TD_TA | Tot. debts/ tot. assets | (long-term debt + current liabilities)/tot. assets | 7 |
| LTD_EP | Long-term debt/shareholder funds | Long-term debt/shareholder funds | |
| NOWC | Net op. work. capital/tot. assets | (current assets-current liabilities)/ tot. assets | |
| TD_EQ | Tot. debt/shareholder funds | (long-term debt + current liabilities)/shareholder funds | |
| **LIQUIDITY** | | | |
| CA_TA | Current assets/tot. assets | Current assets/tot. assets | |
| CR | Current ratio | Current asset/current liabilities | |
| DR | Debt ratio | Total liabilities/tot. assets | |
| WC_TA | Working capital/total assets | Working capital/tot. assets | |
| CASH_CL | Cash/current liability | Cash/current liability | 10 |
| CASH_TA | Cash/tot. assets | Cash/tot. assets | |
| CL_TA | Current liability/tot. assets | Current liability/ tot. assets | |
| TLTA* | One if total liabilities exceeds tot. assets, zero otherwise | One if total liabilities exceeds tot. assets, zero otherwise | |
| CASH_CA | Cash/current assets | Cash/current assets | |
| CF_CL | Cash flow/current liabilities | Cash flow/current liabilities | |
| **SOLVENCY** | | | |
| FE_EBITDA | Financial expenses/EBITDA | Financial expenses/EBITDA | |
| FE_NI | Financial expenses/net income | Financial expenses/net income | 3 |
| FE_TA | Financial expenses/tot. assets | Financial expenses/tot. assets | |
| **TURNOVER** | | | |
| CL_TS | Current liabilities/tot. sales | Current liabilities/tot. sales | |
| CA_TS | Current assets/tot. sales | Current assets / tot. sales | 4 |
| *NAT | *Net asset turnover | *Net sales/tot. assets | |
| WC_TS | Work. Capital/tot. sales | Work. Capital/tot. sales | |
| **ACTIVITY/GROWTH** | | | |
| CF_NS | Cash flow/sales | Cash flow/sales | |
| GROW_TA | Growth ratio of tot. assets | (tot assets/tot. assets t-1)-1 | 4 |
| EBITDA_TS | EBITDA/tot. sales | EBITDA/tot. sales | |
| *NI_GROW | *Net income growth | *($NI_t - NI_{t-1}$)/($|NI_t|+ |NI_{t-1}|$), $NI_t$ : latest net income | |
| **OTHERS** | | | |
| *ORPE | *Operating revenue per employee | *Operating revenue/n.employee | |
| TA | Tot. assets | Tot. assets | 3 |
| SALES | Tot. sales revenue | Tot. sales revenue | |
| **Total FRs** | | | 42 |

Moreover, for each firm, financial data have been collected for up to four years prior the financial distress occurred due to limited data availability on Amadeus Database and, for the sake of simplification in the final results, they have been indicated with year-1, year-2, year-3, year-4. For instance, for a firm that faced financial distress in 2014, the collected financial data span the period 2013-2010 in which 2013 represents the year before its financial distress (year-1), corresponding also to the last year of available information on Amadeus database, and years 2010-2011-2012 represent respectively the year-2, year-3 and year-4 before the company's financial distress. Since the last available data cover a period between 2013 and 2018, the current sample actually covers the period 2013-2018.

Thus, each variable of Table 2 has been considered throughout four years' time span and their selection, implemented only on the training sample, has been performed with the following stages:

(1) A discriminatory power analysis of the 37 FRs through the information value for the four years considered;

(2) The t-test t has been applied on the selected variables of the previous stage;

(3) A correlation analysis has been also performed to eliminate the issue of overlapping information measuring the same characteristics.

4.2. INDEPENDENT VARIABLES SELECTION: INFORMATION VALUE, T-TEST AND CORRELATION ANALYSIS

The sample identification of the variables over the four years considered with the highest explanatory relationship with the credit risk is composed of the three aforementioned steps that we discuss in detail in this section.

**Stage 1.** Information Value ($IV$) has been often used in credit scoring model as benchmark value to distinguish variables with no or weak predictive power, useless for credit risk modelling, from those with medium or high predictive power, decisive in increasing the accuracy of the final model (see Yap et al., 2011 and Nikolic et al., 2013 for its application).

Information Value is computed according to the following formula:

$$IV_i = \sum_{j=1}^{m} \left( (active_j - inactive_j) \cdot WOE_j \right), \qquad (9)$$

where $IV_i$ is the information value of variable $i$ under consideration, $m$ is the total number of companies in the sample, $active_j$ and $inactive_j$ represent respectively the proportion of active and inactive companies for the variable $i$ over $m$, $j$ is the index relative to the company to evaluate and $WOE$ is the weight of evidence, calculated with the formula:

$$WOE_j = ln\left(\frac{active_j}{inactive_j}\right). \qquad (10)$$

Moreover, in order to determine if the predictive power of independent variables is poor, medium or high with respect to company's creditworthiness, the thresholds values, as determined by Siddipi (2012), have been computed (see Table 3).

Table 3. The predictive value of $IV$ according to Siddipi (2012) interpretation.

| Predictive value | $IV$ |
|---|---|
| useless for prediction | <0,05 |
| weak predictor | 0,05<$IV$<0,01 |
| medium predictor | 0,01<$IV$<0,25 |
| strong predictor | 0,25<$IV$<0,50 |
| suspicious or too good to be true | $IV$>0,50 |

With respect to the previous analysis, three variables have been eliminated from information value analysis because of their null or weak predictive power ($IV < 0.1$) in financial distress modelling in at least three years, namely: TD_EQ, CL_TA and FE_TA; otherwise variables in which the information value is predictive in at least two or three years, have been retained for stage 2, together with those variables in which *IV* is predictive in all the considered years.

**Stage 2**. In this step, another useful discriminatory power indicator, the t-test, has been applied on the remaining 34 variables obtained at stage 1, in order to analyze which variables well discriminate on average between failed and not-failed companies. A p-value less than 10% has been considered as confidence interval to define high predictive variables in failure prediction model; otherwise, predictors with a not significant p-value for at least three years have been removed for a further analysis. In this stage, it results that only the following eight variables have been selected: ROA, EBITDA_TA and CF_TA regarding the profitability dimension; EQ_RATIO and TD_TA the financial structure; CA_TA and DR the liquidity condition and CA_TS the turnover aspect.

From stage 2, it is worthy to notice that solvency and activity categories have not predictive power in determining the failure of companies operating in the energy sector as well as the size variables, introduced in this analysis to consider the difference among companies in terms of dimensions. This last result contradicts some credit scoring studies conducted in other sectors, suggesting the significant impact of firm's size on the future companies' probability to fail.

**Stage 3.** Finally, a pairwise correlation analysis has been implemented on the eight variables selected in the previous step for each year of observation (year-1, year-2, year-3 year-4), to eliminate the potential issue of overlapping information leading to the high overfitting on training sample and low performance on test sample. A correlation coefficient greater or equal than |0.5| suggests a high correlation strength between each pair of variables. Table 4 presents the results of the correlation analysis from which it is observed that the CF_TA and DR are highly correlated with at least two other variables over all the considered years.

Table 4. Pearson correlation coefficients among financial variables selected in stage 2 for all periods considered. Source: Statistical Software Stata

| | ROA | EBITDA_TA | CF_TA | EQ_RATIO | TD_TA | CA_TA | DR | CA_TS |
|---|---|---|---|---|---|---|---|---|
| **Year-1** | | | | | | | | |
| ROA | 1 | | | | | | | |
| EBITDA_TA | -0.1373 | 1 | | | | | | |
| CF_TA | 0.6866* | -0.1062 | 1 | | | | | |
| EQ_RATIO | 0.4342 | 0.0957 | 0.5831* | 1 | | | | |
| TD_TA | -0.4037 | -0.0832 | -0.5880* | -0.8448* | 1 | | | |
| CA_TA | -0.2821 | 0.0911 | -0.4155 | -0.3310 | 0.3829 | 1 | | |
| DR | -0.4342 | -0.0957 | -0.5831* | -1.0000* | 0.8448* | 0.3310 | 1 | |
| CA_TS | -0.1349 | -0.0509 | -0.0930 | -0.0541 | 0.0769 | -0.0138 | 0.0541 | 1 |
| **Year-2** | | | | | | | | |
| ROA | 1 | | | | | | | |
| EBITDA_TA | 0.5359* | 1 | | | | | | |
| CF_TA | 0.6122* | 0.9467* | 1 | | | | | |
| EQ_RATIO | 0.2671 | 0.4566 | 0.5365* | 1 | | | | |
| TD_TA | -0.2010 | -0.1583 | -0.1930 | -0.6910* | 1 | | | |
| CA_TA | -0.0634 | -0.2232 | -0.2177 | -0.2874 | 0.3823 | 1 | | |
| DR | -0.2671 | -0.4566 | -0.5365* | -1.0000* | 0.6910* | 0.2874 | 1 | |
| CA_TS | -0.0765 | -0.0976 | -0.0736 | -0.0064 | 0.0417 | -0.0030 | 0.0064 | 1 |
| **Year-3** | | | | | | | | |
| ROA | 1 | | | | | | | |
| EBITDA_TA | 0.7513* | 1 | | | | | | |
| CF_TA | 0.4590 | 0.3496 | 1 | | | | | |
| EQ_RATIO | 0.3317 | 0.2121 | 0.1847 | 1 | | | | |
| TD_TA | -0.2110 | -0.2373 | -0.1039 | -0.7014* | 1 | | | |
| CA_TA | -0.0261 | -0.2426 | 0.0565 | -0.2217 | 0.3151 | 1 | | |
| DR | -0.3317 | -0.2121 | -0.1847 | -1.0000* | 0.7014* | 0.2217 | 1 | |
| CA_TS | -0.0853 | -0.2223 | -0.0859 | -0.0931 | 0.0329 | -0.0225 | 0.0931 | 1 |
| **Year-4** | | | | | | | | |
| ROA | 1 | | | | | | | |
| EBITDA_TA | 0.7554* | 1 | | | | | | |
| CF_TA | 0.9081* | 0.8129* | 1 | | | | | |
| EQ_RATIO | 0.2990 | 0.1106 | 0.3160 | 1 | | | | |
| TD_TA | -0.1754 | -0.1331 | -0.2141 | -0.6974* | 1 | | | |
| CA_TA | 0.0763 | -0.1142 | -0.1651 | -0.1774 | 0.2130 | 1 | | |
| DR | -0.2990 | -0.1106 | -0.3160 | -1.0000* | 0.6974* | 0.1774 | 1 | |
| CA_TS | -0.1677 | -0.2333 | -0.1812 | -0.1117 | 0.0738 | -0.0245 | 0.1117 | 1 |

* correlation coefficient ≥ |0.50|

CF_TA and DR have been removed from further considerations in the analysis to avoid multi-collinearity problems with the development of M.H.DIS model.

Thus, after performing the examined three step procedure, six Financial Ratios have been retained covering different aspects of firms' main features related to profitability (ROA and EBITDA_TA), financial structure (EQ_RATIO and TD_TA), liquidity (CA_TA) and turnover (CA_TS).

Table 2A in the Appendix summarizes the results obtained from the previous steps.

In what follows, we apply the M.H.DIS method using the selected variables (summarized in Table 5) as evaluation criteria.

Table 5. List of the six financial variables selected through the three steps procedure, Authors' elaboration

| Acronym | Variables | Category | Preference direction |
|---|---|---|---|
| ROA | ROA | profitability | max |
| EBITDA_TA | EBITDA/tot. assets | profitability | max |
| EQ_RATIO | Equity ratio | financial structure | max |
| TD_TA | Tot. debts/tot. assets | financial structure | min |
| CA_TS | Current assets/tot. sales | turnover | min |
| CA_TA | Current assets/tot. assets | liquidity | max |

## 5. M.H.DIS MODEL DEVELOPMENT AND MAIN RESULTS

The M.H.DIS model proposed by Zopounidis and Doumpos (2000) has been developed through the following steps.

Initially, a performance matrix has been built in order to organize the dataset in alternatives and criteria; the companies of the full sample constitute alternatives, whereas the six financial ratios selected with the explained three steps procedure, are the criteria under which alternatives are evaluated.

Then criteria with a non-increasing preference direction, such as TD_TA and CA_TS, have been aligned to the criteria with an increasing preference direction, by multiplying the evaluation of alternatives with respect to these criteria for $-1$, and possible outliers have been smoothed through a data trimmed procedure. In this regard, outliers have been identified with the Interquartile Range method (IRQ), by verifying one of the following inequalities (Cinelli, 2017):

$$g_i(a_j) < Q_1 - 1.5(Q_3 - Q_1) \quad \text{or} \quad g_i(a_j) > Q_1 + 1.5(Q_3 - Q_1). \tag{11}$$

Once data have been trimmed to the maximum or minimum values that are not outliers, a five-fold cross validation has been applied to the full sample for year-1. Thus for year-1, the training set of each fold has been used for model development and the test set of each fold has been used for validation purposes.

Moreover, because of the lack for the suitable synthetic judgments released by credit rating agencies (CRAs) for unlisted European energy companies, in this study we consider the classification provided by Amadeus database annually, that sorts companies into two categories: active and inactive ones.

Because of this two-class classification, the M.H.DIS method provides a pair of utility functions in one stage.

In the first stage, the companies belonging to $C_1$ are distinguished from the remaining firms of the other class and two additive utility functions are built: the first one ($U_1$) for $C_1$ and the second one ($U_{\sim 1}$) for the riskier class than $C_1$ (i.e. $C_2$). If the global utility function on the first class is greater than the second utility function ($U_1 > U_{\sim 1}$), then the company is classified into $C_1$; otherwise it belongs to $C_2$.

Table 6 presents the set of weights of the financial ratios in the two utility functions computed. The results indicate that EBITDA_TA and CA_TA are the most significant criteria in discriminating companies of class $C_1$ from companies of class $C_2$.

Table 6. Financial ratios weights in the utility function developed through M.H.DIS model.

| FRs | h1i(%) | h~1i(%) |
|---|---|---|
| ROA | 13.57 | 50.54 |
| EBITDA_TA | 42.79 | 4.69 |
| EQ_RATIO | 8.21 | 2.22 |
| TD_TA | 10.69 | 13.34 |
| CA_TA | 23.55 | 7.29 |
| CA_TS | 1.17 | 21.88 |

Finally, Table 7 displays the classification results of the discriminating model applied for year-1, year-2, year-3 and year-4 on the average of five folds assessed respectively for training and test set. More specifically, Table 1 shows the classification results in terms of companies belonging to each class, correctly or incorrectly predicted by the model. Whereas

Table 2 shows the results in terms of average and overall accuracy rate.

It is important to point out that according to the confusion matrix it is possible to provide the following definitions:

- <u>True Positive (TP)</u>: the number of correctly classified companies belonging to class $C_1$; i.e. companies that according to the Amadeus classification belong to $C_1$ and are classified by the model to the same class $C_1$;
- <u>True Negative (TN)</u>: the number of correctly classified companies belonging to class $C_2$, i.e. companies that according to the Amadeus classification belong to $C_2$ and are classified by the model to the same class $C_2$;
- <u>False Positive (FP)</u>: the number of active companies misclassified as inactive (Type I error); i.e. companies that according to the Amadeus classification belong to $C_1$ and are classified by the model to $C_2$;
- <u>False Negative (FN)</u>: the number of inactive companies misclassified as active (Type II error), i.e. companies that according to the Amadeus classification belong to $C_2$ and are classified by the model to $C_1$.

Moreover, TP and TN, i.e. the companies correctly classified by the model, are located in the main diagonal of the confusion matrix; while FP and FN, i.e. the companies misclassified by the model, are located outside the main diagonal of the matrix. To be more precise, in Table 1 we have denoted the aforementioned acronyms with apices.

The main results of Table 1 suggest that on average the M.H.DIS model developed for year-1 estimates correctly 75 companies on training set $(41\ TP + 35\ TN)$, and 15 companies on test set $(8\ TP + 7\ TN)$. Similarly, for year-2 on average the model estimates correctly 73 companies on training set ($37\ TP + 36\ TN$), and 15 companies on test set ($8\ TP + 7\ TN$). As expected, the number of companies correctly predicted by the model decreases in year-3 and year-4 in both training and test set.

Moreover, the results of the model can be read with respect to the accuracy rate. The two most important measures are: the overall and the average accuracy rate.

- <u>Average accuracy rate (ACA)</u>: is the average of each accuracy per class. It is computed as $\left(\frac{TP}{TP+FP} + \frac{TN}{TN+FN}\right)/classes$;

- <u>Overall accuracy rate (OCA)</u>: is the number of correctly predicted companies over the total companies to predict. ItKosmi is computed as $\frac{TP+TN}{TP+FP+TN+FN}$.

Hence, from

Table 2 it is worth to note that on average the model developed for year-1 estimates correctly the 83.36% (OCA) of companies of training set and less than the 70% of companies for the test set (OCA 68.18%). This trend decreases with years, by reaching an overall accuracy rate of 56.36% for year-4. This result confirms what we expect, namely that on average the model estimates better companies where data are collected more recently than where data are collected from the past. The reason is quite intuitive since the predefined classification provided by Amadeus is more likelihood to be affected by the last balance sheet data (year-1) than the previous ones (year-2, year-3, year-4).

Furthermore, the overall classification accuracy for year-1, in both training and test sample, is not quite satisfactory compared to other studies on M.H.DIS model applications. Indeed here, on average the M.H.DIS model classifies correctly 83.26% companies of basic sample and 68.18% of holdout sample against the percentage range of [100, 93.18] and [75.11, 69] respectively for training and test set of other similar studies in other sectors (Kosmidou et al., 2002; Doumpos and Zopounidis, 1999; Kosmidou et al., 2004).

Table 1. Classification results of M.H.DIS model in terms of companies belonging to each class for year-1, year-2, year-3, year-4 (average over 5-fold cross-validation for training and test set). Source: Matlab Software

| PREDEFINED CLASSIFICATION | M.H.DIS MODEL ESTIMATED WITH AMADEUS CLASSIFICATION ||||||||||||
| | TRAINING SET ||||||||||||
| | Companies belonging to each class ||||||||||||
| | Year-1 ||| Year-2 ||| Year-3 ||| Year-4 |||
| | $C_1$ | $C_2$ | Tot | $C_1$ | $C_2$ | Tot | $C_1$ | $C_2$ | Tot | $C_1$ | $C_2$ | Tot |
| $C_1$ | 41$^{(TP)}$ | 5$^{(FP)}$ | 46 | 37$^{(TP)}$ | 9$^{(FP)}$ | 46 | 40$^{(TP)}$ | 6$^{(FP)}$ | 46 | 37$^{(TP)}$ | 9$^{(FP)}$ | 46 |
| $C_2$ | 11$^{(FN)}$ | 35$^{(TN)}$ | 46 | 10$^{(FN)}$ | 36$^{(TN)}$ | 46 | 16$^{(FN)}$ | 30$^{(TN)}$ | 46 | 13$^{(FN)}$ | 33$^{(TN)}$ | 46 |
| Tot | | | 92 | | | 92 | | | 92 | | | 92 |
| | TEST SET ||||||||||||
| PREDEFINED CLASSIFICATION | Companies belonging to each class ||||||||||||
| | Year-1 ||| Year-2 ||| Year-3 ||| Year-4 |||

|       | $C_1$ | $C_2$ | Tot | $C_1$ | $C_2$ | Tot | $C_1$ | $C_2$ | Tot | $C_1$ | $C_2$ | Tot |
|-------|-------|-------|-----|-------|-------|-----|-------|-------|-----|-------|-------|-----|
| $C_1$ | 8(TP) | 3(FP) | 11  | 8(TP) | 3(FP) | 11  | 8(TP) | 3(FP) | 11  | 7(TP) | 4(FP) | 11  |
| $C_2$ | 4(FN) | 7(TN) | 11  | 4(FN) | 7(TN) | 11  | 6(FN) | 5(TN) | 11  | 6(FN) | 5(TN) | 11  |
| Tot   |       |       | 22  |       |       | 22  |       |       | 22  |       |       | 22  |

(TP)-True Positive; (TN)-True Negative; (FN)-False Negative; (FP)-False Positive

Table 2. Classification results of M.H.DIS model in terms of Average and Overall accuracy rate for year-1, year-2, year-3, year-4 (average over 5-fold cross-validation for training and test set). Source: Matlab Software

| M.H.DIS MODEL ESTIMATED WITH AMADEUS CLASSIFICATION | | | | | | | | |
|---|---|---|---|---|---|---|---|---|
| TRAINING SET | | | | | | | | |
| PREDEFINED CLASSIFICATION | Accuracy (%) | | | | | | | |
|  | Year-1 | | Year-2 | | Year-3 | | Year-4 | |
|  | $C_1$ | $C_2$ | $C_1$ | $C_2$ | $C_1$ | $C_2$ | $C_1$ | $C_2$ |
| $C_1$ | 89.95 | 10.05 | 80.37 | 19.63 | 87.43 | 12,57 | 81.22 | 18.77 |
| $C_2$ | 23.36 | 76.64 | 21.21 | 78.79 | 34.57 | 65.43 | 29.46 | 70.53 |
| Average accuracy (%) | 83.29 | | 79.58 | | 76.43 | | 75.87 | |
| Overall accuracy (%) | 83.26 | | 79.56 | | 76.31 | | 75.86 | |
| TEST SET | | | | | | | | |
| PREDEFINED CLASSIFICATION | Accuracy (%) | | | | | | | |
|  | Year-1 | | Year-2 | | Year-3 | | Year-4 | |
|  | $C_1$ | $C_2$ | $C_1$ | $C_2$ | $C_1$ | $C_2$ | $C_1$ | $C_2$ |
| $C_1$ | 75.65 | 24.35 | 68.44 | 31.56 | 72.19 | 27.81 | 64.70 | 35.29 |
| $C_2$ | 38.12 | 61.88 | 35.83 | 64.17 | 57.35 | 42.65 | 52.09 | 47.90 |
| Average accuracy (%) | 68.77 | | 66.31 | | 57.42 | | 56.30 | |
| Overall accuracy (%) | 68.18 | | 66.36 | | 57.27 | | 56.36 | |

It has to be pointed out that this not fully satisfactory result of M.H.DIS model might depend on the rough classification provided by Amadeus database rather than the one provided by credit rating agencies (CRAs). Indeed these latter usually provide an objective synthetic credit rating for each company that is surely more reliable than the one provided by Amadeus for applying the M.H.DIS model. Unfortunately, this type of information is not provided here since the original database is composed by unlisted companies. Hence, the PROMETHEE II model, an acknowledged MCDA model, has been further implemented on the same dataset, to realize whether the original balanced classification provided by Amadeus database, could vary with the application of this model.

## 6. PROMETHEE BASED CLASSIFICATION

To overcome the issue that the M.H.DIS model also in the most recent year-1 does not achieve highly satisfactory results in the holdout sample and for comparison purposes, another two-class assignment of the energy companies in the considered sample has been built on the basis of a well-known multi-criteria decision aid model, the PROMETHEE II method.

PROMETHEE II method has been applied in this study, to provide a benchmark sorting procedure on which to compare the classification provided by Amadeus database. Although there are other PROMETHEE methods specifically developed for solving sorting problems, such as PROMETHEE

TRI and PROMSORT, the PROMETHEE II method has been selected here for several practical reasons.

From one side PROMETHEE TRI and PROMSORT present some important limits in terms of inputs needed for implementing the model. Both models indeed, require two important inputs to introduce: the reference alternatives and the limit profiles. Thus, a priori definition of these elements generates great constraints for practical applications, since it needs the assessment of industry experts. Although in PROMSORT model, the issue to have pre-defined reference alternatives can be can be tackled with PROMETHEE I method (Araz and Ozkarahan, 2007), in PROMETHEE TRI such issue is still present (Figueira et al., 2004). Moreover, the PROMETHEE TRI model presents also the disadvantages to use only single criterion net flows as inputs rather than outranking relation between alternatives, giving back also not perfectly ordered categories.

From the other side PROMETHEE II method presents some advantages in comparison to the aforementioned models, such as its easiness of implementation, its wide practical applications also to credit scoring models (Hu and Chen, 2011; Mousavi and Lin, 2020 ) and its feature to provide a complete ranking of alternatives.

To implement PROMETHEE II method, firstly, from the six financial variables selected in previous stage, we have built all the subsets composed of three criteria i.e. $C_{6,3} = 20$. Then, for each subset we have considered its complement, forming a pair of disjoint subsets denoted as follows:

$$\mathcal{P} = (F, F^c) \ \forall F \subset G \text{ formed of three variables.} \qquad (12)$$

However, the twenty pairs of subsets composed of three criteria (denoted with $\mathcal{P}$) have been reduced to eight, according to the following rules:

(1) If $\mathcal{P}$ is composed of subsets of criteria with a high pairwise correlation ( $> |0.5|$) then such pair has been removed. For example, in our sample the couples of criteria ROA and EBITDA_TA, EQ_RATIO and TD_TA show a high pairwise correlation (see Table 4);

(2) If $\mathcal{P}$ is composed of subsets of at least two criteria belonging to the same dimension, i.e. profitability, financial structure, liquidity or turnover, then it has been eliminated. Since the considered dimensions are four, according to the above rule every subset (and its complement) related to each $\mathcal{P}$ is composed of three criteria representing three different company's aspects.

Applying these rules, eight of the twenty pairs have been retained. To achieve clarity in notation, we denote each of the eight pairs with a numerical label, i.e. $\mathcal{P}_h = (F_h, F_h^c)$ with $h \in \{1, 2, \cdots, 8\}$.

Table 9 displays the eight pairs considered listing the three criteria for each subset and its complement.

The three criteria (or financial variables) belonging to each set $F_h \in \mathcal{P}_h$ have been used in PROMETHEE II as evaluation criteria on which companies' classification is based, whereas the remaining three ones belonging to its relative complement $F_h^c \in \mathcal{P}_h$, have been employed as evaluation criteria of M.H.DIS model development with respect to both the AMADEUS and PROMETHEE  based classification.

Table 9. Pairs ($\mathcal{P}$) considered in our analysis, Authors' elaboration.

|  | Criteria employed | |
| --- | --- | --- |
|  | $F_h$<br>PROMETHEE II Classification | $F_h^c$<br>M.H.DIS Model development |
| $\mathcal{P}_1$ | ROA<br>EQ_RATIO<br>CA_TS | EBITDA_TA<br>TD_TA<br>CA_TA |
| $\mathcal{P}_2$ | ROA<br>EQ_RATIO<br>CA_TA | EBITDA_TA<br>TD_TA<br>CA_TS |
| $\mathcal{P}_3$ | ROA<br>TD_TA<br>CA_TS | EBITDA_TA<br>TD_TA<br>CA_TA |
| $\mathcal{P}_4$ | ROA<br>TD_TA<br>CA_TA | EBITDA_TA<br>EQ_RATIO<br>CA_TS |
| $\mathcal{P}_5$ | EBITDA_TA<br>EQ_RATIO<br>CA_TS | ROA<br>TD_TA<br>CA_TA |
| $\mathcal{P}_6$ | EBITDA_TA<br>EQ_RATIO<br>CA_TA | ROA<br>TD_TA<br>CA_TS |
| $\mathcal{P}_7$ | EBITDA_TA<br>TD_TA<br>CA_TS | ROA<br>EQ_RATIO<br>CA_TA |
| $\mathcal{P}_8$ | EBITDA_TA<br>TD_TA<br>CA_TA | ROA<br>EQ_RATIO<br>CA_TS |

This procedure has several advantages. Firstly, only those criteria that well discriminate companies' dimensions are used to sort energy firms into classes, giving an increasing consistency to the PROMETHEE based companies' classification than the one collected from the AMADEUS database. Secondly, by considering $F_h^c$, relative to each $\mathcal{P}_h$, it is possible to develop the M.H.DIS model on the PROMETHEE based classification, which represents a benchmark to compare the classification performances of AMADEUS based classification in terms of overall accuracy.

Thirdly, taking into account all the considered subsets of three criteria ($F_h^c$ with $h \in \{1, 2, \cdots, 8\}$), it is highlighted how the overall accuracy varies according to the pair considered, acting as robustness check if M.H.DIS model built with PROMETHEE classification achieves higher results than the ones obtained with AMADEUS classification for most of the pairs.

PROMETHEE II, being founded on six types of preference functions, can potentially yield a different companies' classification according to the preference function used; if the obtained classification does not vary very much, then the model is quite consistent regardless the preference function used, representing a further element of robustness.

Finally, by simulating different scenarios for criteria weights, we performed a robustness analysis also with respect to the assessment of the PROMETHEE evaluation of each company and consequently on its assignment to a class.

### 7. RESULTS AND DISCUSSIONS OF M.H.DIS MODEL DEVELOPED ON THE PROMETHEE BASED CLASSIFICATION

In this section, we discuss the results of the M.H.DIS model developed respectively with AMADEUS and PROMETHEE based classification, which we will build before applying the multicriteria discrimination model. To deal with this aim, once data have been trimmed and criteria with a non-increasing preference direction have been aligned to the ones with an increasing preference function (Section 5), PROMETHEE II method has been applied with respect to the three criteria belonging to $F_h$ of each $\mathcal{P}_h$ (Table 9, 2$^{nd}$ column) by considering the six type preference functions described in Section 2.2. For each Preference Function (PF), alternative ($a_j$) and set of three criteria ($F_h \in \mathcal{P}_h$), we obtain a net flow $\Phi(a_j) \in [-1,1]$ that allows to rank alternatives from the best to the worst. In order to classify companies into two categories, the healthiest ($C_1$) and the riskiest class ($C_2$), we employ the median of the net flow of the all alternatives as a cut-off limiting the two classes.

In this framework, the six preference functions of PROMETHEE II have been considered for each set of criteria $F_h \in \mathcal{P}_h$ with $h \in \{1, 2, \cdots, 8\}$. Thus, we get in total forty-eight classifications of companies obtained multiplying the eight subsets ($F_h$) considered (see Table 9) by six type preference functions. Hence, the achieved classifications might differ each other according to the preference function and the set of criteria $F_h$ considered. However, it has to be pointed out that the majority of preference functions (in at least four of the six type functions) classify companies in the same manner. In this regard, Table 10 shows the classification of companies into the healthiest and riskiest class according to the majority of the preference functions, with their relative and cumulative frequency for each combination.

Table 10. Companies' classification according to the most preference functions employed in PROMETHEE II. Authors' elaboration.

|  | Classification of companies according to the majority of preference functions | | | |
|---|---|---|---|---|
|  | Class | Number of companies | Relative frequency (%) | Cumulative frequency (%) |
| $F_1^c$ | 1 | 72 | 63.16 | - |
|  | 2 | 40 | 35.09 | 63.16 |
|  | not perfectly determined by most of PF | 2 | 1.75 | 98.25 |
|  | total | 114 | 100 | 100 |
| $F_2^c$ | 1 | 45 | 39.47 | - |
|  | 2 | 65 | 57.01 | 39.47 |
|  | not perfectly determined by most of PF | 4 | 3.50 | 96.49 |
|  | total | 114 | 100 | 100 |
| $F_3^c$ | 1 | 82 | 71.92 | - |
|  | 2 | 30 | 26.31 | 71.92 |
|  | not perfectly determined by most of PF | 2 | 1.75 | 98.24 |
|  | total | 114 | 100 | 100 |
| $F_4^c$ | 1 | 58 | 50.87 | - |
|  | 2 | 55 | 48.24 | 50.87 |
|  | not perfectly determined by most of PF | 1 | 0.87 | 99.12 |
|  | total | 114 | 100 | 100 |
| $F_5^c$ | 1 | 83 | 72.8 | - |
|  | 2 | 29 | 25.43 | 72.80 |
|  | not perfectly determined by most of PF | 2 | 1.75 | 98.24 |
|  | total | 114 | 100 | 100 |
| $F_6^c$ | 1 | 70 | 61.40 | - |
|  | 2 | 43 | 37.71 | 61.40 |
|  | not perfectly determined by most of PF | 1 | 0.87 | 99.12 |
|  | total | 114 | 100 | 100 |
| $F_7^c$ | 1 | 90 | 78.94 | - |
|  | 2 | 22 | 19.29 | 78.94 |
|  | not perfectly determined by most of PF | 2 | 1.75 | 98.24 |
|  | total | 114 | 100 | 100 |
| $F_8^c$ | 1 | 80 | 70.17 | - |
|  | 2 | 30 | 26.31 | 70.17 |
|  | not perfectly determined by most of PF | 4 | 3.50 | 96.49 |
|  | total | 114 | 100 | 100 |

Two main elements can be observed from Table 10:

(1) the significant difference between the classification obtained with the PROMETHEE method and the one provided by AMADEUS database;
(2) the robustness of the PROMETHEE II method to sort companies.

With regard to the first point, PROMETHEE model classifies, in six of the eight combinations ($F_h^c$ with $h = 1, 3, 5, 6, 7$ and $8$), most of companies as healthiest with a relative frequency that ranges between 61.40% and 78.94%; on the contrary AMADEUS based classification is equally distributed among the two classes (see Table 1).

With regard to the second point, PROMETHEE based method represents a robust tool to sort companies, since in each combination the majority of the preference functions (in at least four of the six type functions) provides a consistent classification regardless of the preference function employed. Moreover, those companies for which most preference functions are not able to determine with a strict preference the membership to healthiest or riskiest class, are limited to very few cases (from one to four companies).

Table 11 presents the main results of M.H.DIS model for year-1, developed respectively for AMADEUS and PROMETHEE based classification. Furthermore, in order to compare the efficiency of the discrimination model on two different rating settings, different performance indicators are needed. Among the most widely applied to assess the performance of credit rating models (Sobehart and Keenan, 2001; Keenan and Sobehart, 1999; Engelmann et al., 2003; Tinoco and Wilson, 2013) there are:

- Cumulative Accuracy Profiles (CAP): is a graphical representation of two CAP curves that help to visualize the global performance of a model to discriminate two groups. However, to plot these curves it is necessary that companies have to be ranked by risk score. Random models display a curve coincident with the main diagonal of the graph; while perfect models show a line steeper to the left and closer to the point (0, 1);
- Sensitivity (SENS): is a measure of how well a model identifies True Positive. It is given by the number of non-defaulted companies evaluated correctly by the model ($TP$), over the total number of non-defaulted companies ($TP + FP$);
- Specificity (SPEC): is a measure of how well a model identifies True Negatives. It is computed by the number of defaulted companies evaluated correctly by the model (TN) over the total number of defaulted companies ($TN + FN$);
- Classification Accuracy (CA): it is a single summary measure that examines whether a company is classified correctly by the model without considering the magnitude of misclassification. It can be distinguished into average and overall accuracy rate (ACA and OCA);
- Receiver Operating Characteristic (ROC): it is a graphical plot similar to the CAP that provides a sketch of rating scores' distribution for active and inactive companies (Fawcett, 2006). However, it presents results that are more intuitive than CAP. The rating model's performance is the better when the ROC curve is steeper to the left and closer to the point (0, 1);
- Area Under the Receiver Operating Characteristic curve (AUROC): it is the summary statistic of the ROC curve and it is a standard measure for the predictive accuracy of the model. It represents the likelihood that an active company will obtain a higher credit score compared to an inactive company, by measuring the area between the curve and the diagonal of the Lorenz curve (Fawcett, 2006). AUROC values range between 0-1. The model assumes a value equal to 0.5 whether it is random or lacks discriminative power; while it takes a value equal to 1 whether it perfectly discriminates among groups. Generally, models takes values between 0.5 and 1;

- Gini Coefficient (GINI): it is widely used to assess the predictive accuracy of training and test set (Altman et al., 2010). It is easy to interpret and compute since it derives from AUROC, but differs for computing the full area below the curve. Hence, following the approach of Altman et al. (2010), it can be computed as $(2 * AUROC) - 1$. A Gini coefficient greater than 0.5 can be considered satisfactory;
- Kolmogorov-Smirnov distance (KSD): it measures the maximum vertical deviation between two cumulative distributions functions. It has been mainly used to evaluate the predictive accuracy of USA rating systems jointly with other performance indicators (Andersen, 2007). Acceptable values of KS range between [20%, 70%]; if values are higher than 70%, the model is too good to be true (Mays, 2004);
- F1_Score: is one of the most used indicators for machine learning applications not only for a binary classification, but also for multiple classification. It is a weighted harmonic mean of Recall and Precision test (Powers, 2015). Recall is the Sensitivity, while Precision is the Specificity.

In this Chapter, the six performance indicators of Table 3 have been selected from the previous list to compare the discriminating performance of M.H.DIS model developed respectively with Amadeus and PROMETHEE based classification. More specifically, following the approach of Doumpos et al. (2016), we have selected only those measures deriving from the main elements of confusion matrix (TP, TN, FP, FN) (Section 5) and endowed of higher computational intelligibility than graphical. Thus, performance indicators such as CAP, RO, KSD have been discarded because of their high graphic evidence; conversely SENS, SPEC, ACA, OCA, AUROC and Gini coefficient have been retained for their high quantitative evidence. Gini coefficient, in particular, has been included among these measures to check the consistency of the other performance indicators involved into the efficiency analysis.

In

Table 4, performance indicators with respect to each preference function, used to develop M.H.DIS model with PROMETHEE-based classification, that are lower than the ones obtained with AMADEUS, are denoted with asterisk (*).

Table 3. Performance indicators used to evaluate the efficiency of M.H.DIS model. Authors' elaboration.

| Performance Indicators | | | | |
|---|---|---|---|---|
| Acronym | Indicator's name | Formula | Value (%) | Pref. direction |
| SENS | Sensitivity | $\frac{TP}{TP + FP}$ | [0-100] | max |
| SPEC | Specificity | $\frac{TN}{TN + FN}$ | [0-100] | max |
| ACA | Average accuracy | $\frac{SENS + SPEC}{classes}$ | [0-100] | max |
| OCA | Overall accuracy | $\frac{TP + TN}{TP + FP + TN + FN}$ | [0-100] | max |
| AUROC | Area under the receiving operating characteristic | $\frac{1}{2} * \left(\frac{TP}{TP + FN} + \frac{TN}{TN + FP}\right)$ | [50-100] | max |
| GINI | Gini coefficient | $(2 * AUROC) - 1$ | [-100; 100] | max |

Table 4. Results of M.H.DIS model for year-1 developed for AMADEUS and PROMETHEE classification (average over 5-fold cross-validation for training and test set).

| | Criteria employed | | TRAINING AND TEST SET | M.H.DIS MODEL YEAR-1 | | | | | | | |
|---|---|---|---|---|---|---|---|---|---|---|---|
| | | | | | | | PROMETHEE CLASSIFICATION | | | | |
| | $F_h$ PROMETHEE II | $F_h^c$ M.H.DIS Model Development | | PERFORMANCE INDICATORS | AMADEUS CLASSIFICATION | REGULAR | U-SHAPE | V-SHAPE | LEVEL | LINEAR | GAUSSIAN |
| $\mathcal{P}_1$ | ROA EQ_RATIO CA_TS | EBITDA_TA TD_TA CA_TA | TRAINING SET | SENS SPEC ACA OCA AUROC GINI | 86.00 66.64 76.32 76.30 77.64 55.28 | 94.79 73.56 84.17 85.00 86.57 73.14 | 94.56 80.31 87.43 88.91 89.34 78.68 | 92.46 81.45 86.95 88.48 87.97 75.94 | 95.50 81.31 88.40 90.43 90.93 81.87 | 86.76 83.74 85.25 86.09 79.99 59.98 | 86.76 83.74 85.25 86.09 79.99 59.98 |
| | | | TEST SET | SENS SPEC ACA OCA AUROC GINI | 70.62 50.77 60.70 60.91 61.01 22.01 | 84.21 65.51 74.86 74.55 75.43 50.86 | 87.64 73.50 80.57 80.91 81.48 62.95 | 87.06 72.78 79.92 80.00 79.96 59.91 | 87.26 72.00 79.63 80.00 80.25 60.50 | 82.93 72.29 77.61 79.09 72.21 44.42 | 82.93 72.29 77.61 79.09 72.21 44.42 |
| $\mathcal{P}_2$ | ROA EQ_RATIO CA_TA | EBITDA_TA TD_TA CA_TS | TRAINING SET | SENS SPEC ACA OCA AUROC GINI | 86.86 68.39 77.63 77.61 78.68 57.36 | 86.42* 74.97 80.69 81.09 81.37 62.73 | 82.33* 73.79 78.06 77.61 77.97* 55.94* | 87.05 72.81 79.93 78.70 79.18 58.37 | 85.65* 68.12* 76.89* 75.00 76.12* 52.25* | 84.54* 71.61 78.07 76.31* 76.12* 52.23* | 87.96 66.90* 77.43* 73.91* 74.78* 49.56* |
| | | | TEST SET | SENS SPEC ACA OCA AUROC GINI | 76.60 61.17 68.89 69.09 69.23 38.47 | 72.59* 66.24 69.42 69.09 70.05 40.10 | 63.07* 63.99 63.53* 62.73* 64.08* 28.16* | 84.27 68.64 76.46 74.55 75.96 51.92 | 70.55* 56.81* 63.68* 62.73* 63.51* 27.02* | 71.31* 57.31* 64.31* 61.82* 63.28* 26.56* | 74.98* 58.14* 66.56* 64.55* 65.80* 31.60* |
| $\mathcal{P}_3$ | ROA TD_TA CA_TS | EBITDA_TA TD_TA CA_TA | TRAINING SET | SENS SPEC ACA OCA AUROC GINI | 79.73 69.19 74.46 74.56 75.38 50.75 | 94.92 83.69 89.31 90.00 90.73 81.46 | 93.36 84.60 88.98 90.00 90.15 80.30 | 86.09 91.78 88.94 87.61 83.66 67.31 | 89.03 92.31 90.67 90.00 87.09 74.17 | 90.14 95.54 92.84 91.09 83.79 67.57 | 90.37 95.99 93.18 91.52 85.83 71.66 |
| | | | TEST SET | SENS SPEC ACA OCA AUROC GINI | 69.34 55.92 62.63 61.82 62.95 25.90 | 87.70 73.53 80.61 81.82 81.72 63.45 | 90.08 65.29 77.68 81.82 83.29 66.57 | 77.41 68.00 72.71 73.64 68.90 37.80 | 77.22 74.79 76.00 75.46 71.41 42.82 | 82.97 79.29 81.13 80.91 73.32 46.64 | 86.60 80.29 83.44 83.80 79.24 58.47 |
| $\mathcal{P}_4$ | ROA TD_TA CA_TA | EBITDA_TA EQ_RATIO CA_TS | TRAINING SET | SENS SPEC ACA OCA AUROC GINI | 86.46 68.87 77.67 77.61 78.51 57.03 | 93.77 80.32 87.05 87.82 88.67 77.35 | 88.63 74.79 81.71 81.52 82.27 64.55 | 85.56* 80.38 82.97 83.04 83.17 66.35 | 83.25* 76.77 80.01 80.00 80.61 61.23 | 68.54* 81.11 74.83* 75.44* 77.27* 54.55* | 67.96* 79.71 73.84* 74.13* 76.45* 52.89* |
| | | | TEST SET | SENS SPEC ACA OCA AUROC GINI | 80.08 52.84 66.46 66.36 67.65 35.30 | 87.65 67.00 77.33 78.18 78.96 57.93 | 81.13 64.18 72.66 71.82 73.40 46.80 | 77.46* 70.99 74.22 74.55 74.66 49.32 | 77.68* 68.62 73.15 73.64 73.26 46.53 | 55.43* 64.38 59.91* 60.00* 60.16* 20.31* | 58.89* 60.51 59.70* 60.00* 59.78* 19.55* |
| $\mathcal{P}_5$ | EBITDA_TA EQ_RATIO CA_TS | ROA TD_TA CA_TA | TRAINING SET | SENS SPEC ACA OCA AUROC GINI | 89.81 60.52 75.16 75.22 78.52 57.05 | 88.12* 84.60 86.36 86.52 86.62 73.25 | 81.17* 95.31 88.24 85.65 83.98 67.97 | 84.33* 90.23 87.28 85.87 81.70 63.40 | 84.33* 90.23 87.28 85.87 81.70 63.40 | 80.24* 87.43 83.83 82.17 78.51* 57.01* | 77.28* 89.41 83.34 80.00 77.10* 54.21* |
| | | | TEST SET | SENS SPEC ACA OCA AUROC GINI | 75.09 48.43 61.76 61.82 63.60 27.20 | 80.47 72.37 76.42 76.36 76.27 52.54 | 75.25 82.43 78.84 77.27 74.48 48.95 | 79.85 82.38 81.12 79.09 76.07 52.15 | 79.85 76.67 78.26 77.27 73.42 46.84 | 78.97 72.14 75.56 73.64 73.75 47.51 | 73.72* 75.00 74.36 71.82 61.93* 23.86* |
| $\mathcal{P}_6$ | EBITDA_TA EQ_RATIO CA_TA | ROA TD_TA CA_TS | TRAINING SET | SENS SPEC ACA OCA AUROC GINI | 88.12 73.08 80.60 80.65 81.62 63.25 | 93.90 81.44 87.67 88.91 89.37 78.74 | 92.05 86.73 89.39 89.78 90.35 80.71 | 84.35* 88.38 86.37 85.87 82.60 65.20 | 86.09* 85.78 85.93 85.87 85.43 70.85 | 85.64* 83.74 84.69 85.00 85.36 70.72 | 84.22* 87.54 85.88 85.43 85.77 71.55 |
| | | | TEST SET | SENS SPEC ACA OCA AUROC GINI | 75.09 56.42 65.76 65.46 66.54 33.07 | 79.35 67.33 73.34 74.55 73.40 46.81 | 78.35 75.56 76.95 76.37 77.06 54.13 | 69.25* 71.91 70.58 70.00 69.24 38.49 | 73.19 65.50 69.35 70.91 67.82 35.65 | 72.32* 70.86 71.59 69.00 70.62 41.23 | 68.30* 66.19 67.25 67.27 66.16* 32.33* |
| $\mathcal{P}_7$ | EBITDA_TA TD_TA CA_TS | ROA EQ_RATIO CA_TA | TRAINING SET | SENS SPEC ACA OCA AUROC GINI | 90.13 57.41 73.77 73.91 78.62 57.25 | 89.81 83.19 86.50 86.96 86.91 73.83 | 82.32* 91.59 86.96 85.22 83.31 66.62 | 86.81 87.93 87.37 87.17 80.67 61.34 | 87.33 86.70 87.01 87.39 80.90 61.81 | 90.82 82.11 86.46 89.35 80.57 61.14 | 90.82 82.11 86.46 89.35 80.57 61.14 |
| | | | TEST SET | SENS SPEC ACA OCA | 76.34 48.69 62.51 61.82 | 78.54 73.27 75.90 75.45 | 76.04* 78.09 77.07 75.45 | 79.67 65.71 72.69 74.54 | 80.99 68.43 74.71 76.36 | 85.53 52.38 68.95 79.09 | 85.53 52.38 68.95 79.09 |
|   |   |   |   | AUROC | 65.82 | 76.56 | 74.30 | 70.26 | 70.80 | 65.40* | 65.40* |
|   |   |   |   | GINI | 31.64 | 53.13 | 48.60 | 40.51 | 41.61 | 30.80* | 30.80* |
|   |   |   |   | SENS | 92.15 | 92.16 | 94.14 | 88.19* | 88.89* | 89.44* | 89.50* |
|   | EBITDA_TA | ROA | TRAINING | SPEC | 65.82 | 72.94 | 78.38 | 84.61 | 86.12 | 92.57 | 93.78 |
|   |   |   | SET | ACA | 78.99 | 82.55 | 86.26 | 86.40 | 87.51 | 91.01 | 91.64 |
|   |   |   |   | OCA | 78.91 | 85.65 | 88.26 | 87.17 | 88.26 | 90.22 | 90.44 |
|   |   |   |   | AUROC | 81.12 | 84.91 | 88.48 | 84.17 | 83.88 | 85.73 | 84.66 |
| $\mathcal{P}_8$ | TD_TA | EQ_RATIO |   | GINI | 62.23 | 69.81 | 76.95 | 68.35 | 67.76 | 71.45 | 69.31 |
|   |   |   |   | SENS | 82.21 | 85.90 | 85.38 | 84.88 | 84.57 | 87.81 | 86.27 |
|   |   |   |   | SPEC | 55.84 | 60.33 | 67.12 | 59.52 | 59.05 | 70.09 | 74.21 |
|   | CA_TA | CA_TS | TEST SET | ACA | 69.03 | 73.12 | 76.25 | 72.20 | 71.81 | 78.95 | 80.24 |
|   |   |   |   | OCA | 69.09 | 77.27 | 78.18 | 78.18 | 79.09 | 83.64 | 82.73 |
|   |   |   |   | AUROC | 71.09 | 74.99 | 78.45 | 73.85 | 75.49 | 78.42 | 79.07 |
|   |   |   |   | GINI | 42.18 | 49.97 | 56.90 | 47.70 | 50.99 | 56.84 | 58.15 |

Data on performance indicators are expressed in percentage.

The results clearly show that the discrimination power of M.H.DIS model developed with PROMETHEE based classification is higher than the one obtained with AMADEUS classification, in most of the combinations of criteria relative to subsets $F_h^c$ with $h = 1, 3, 5, 6, 7$ and $8$ for both training and test sample. In these combinations the specificity, the average and the overall accuracy rate of M.H.DIS with PROMETHEE based classification are strictly higher than the ones obtained with AMADEUS classification, regardless the preference function used to develop the PROMETHEE model. Instead, the combinations of criteria relative to subsets $F_h^c$ with $h = 2$ and $4$ do not achieve the same high-performance results especially with regard to the test set.

However, it is observed that in no combination, the performance indicators relative to the AMADEUS classification achieve the maximum value as with PROMETHEE-based classification, but they take an intermediate value within the range of possible six values obtained according to the different preference functions employed in PROMETHEE classification. In other words, it exists at least one or more preference functions also for the less performing combinations 2 and 4, in which the M.H.DIS model performed with PROMETHEE-based classification, gives an accuracy rate that is higher than the one achieved with AMADEUS classification.

Moreover, the combinations of criteria relative to subsets $F_h^c$ with $h = 1, 3, 5, 6, 7$ and $8$ with the highest accuracy rate present some important common features:

- the PROMETHEE-based classification, on which M.H.DIS is developed, is not equally distributed among the two classes, but is more concentrated on the healthiest companies, with a relative frequency ranging between 61.40% and 78.94%;
- the M.H.DIS model is developed by using at least two of the financial variables with a higher weight in discriminating between categories (see Table 5) such as: EBITDA_TA, CA_TA and ROA, with the only exception of the combination of criteria referred to $F_6^c$;
- three of the performance indicators, i.e. Sensitivity, Auroc and Gini coefficient, computed for the M.H.DIS model developed with PROMETHEE based classification achieve the lowest results whenever the preference function employed is more complex such as the level, the linear and Gaussian criterion (see PROMETHEE classification in the combinations of criteria relative to subsets $F_h^c$ with $h = 1, 3, 5, 6, 7$ and 8).

On the contrary, combinations of criteria relative to subsets $F_2^c$ and $F_4^c$ achieve a quite limited accuracy rate and share the following common aspects:

- a PROMETHEE-based classification more equally distributed among categories of companies, such as the AMADEUS classification, with a relative frequency ranging between 39.47% and 50.87%;
- the M.H.DIS model is developed on financial variables with a lower weight in discriminating between categories (see Table 5) such as: EQ_RATIO, TD_TA and CA_TS;

- most of the performance indicators (including also the average and the overall accuracy rate) computed for the M.H.DIS model developed with PROMETHEE based classification achieve the lowest results whenever the preference function employed is more complex such as for the level, the linear and Gaussian ones.

Finally, to prove the robustness of M.H.DIS model developed in PROMETHEE-based classification with respect to the AMADEUS one, M.H.DIS model developed for year-1 has been also applied to the training and test sample for year-2, year-3, year-4.

For the sake of simplification, in Table 12 the performances of two models have been presented in terms of average and overall accuracy rate. Moreover, only the minimum and maximum values of M.H.DIS model with PROMETHEE based classification, attained considering the six-preference functions on the average of 5-fold cross-validation, have been displayed.

Table 12. Results of M.H.DIS model for year-1, year-2, year-3, year-4, developed for AMADEUS and PROMETHEE-based classification (average over 5-fold cross-validation for training and test set).

| | | M.H.DIS MODEL | | | | | | | | | | |
|---|---|---|---|---|---|---|---|---|---|---|---|---|
| | | YEAR-1 | | | YEAR-2 | | | YEAR-3 | | | YEAR-4 | | |
| | TRAINING AND TEST SET | PERF. INDICATORS | AMADEUS CLASSIF. | PROMETHEE CLASSIF. | | AMADEUS CLASSIF. | PROMETHEE CLASSIF. | | AMADEUS CLASSIF. | PROMETHEE CLASSIF. | | AMADEUS CLASSIF. | PROMETHEE CLASSIF. | |
| | | | | MIN | MAX | | MIN | MAX | | MIN | MAX | | MIN | MAX |
| $\mathcal{P}_1$ | TRAINING SET | ACA | 76.32 | 84.17 | 88.40 | 73.23 | 79.45 | 84.00 | 69.79 | 76.80 | 82.61 | 69.07 | 76.65 | 84.59 |
| | | OCA | 76.30 | 85.00 | 90.43 | 73.26 | 79.35 | 86.52 | 69.78 | 79.35 | 83.04 | 69.13 | 72.17 | 85.22 |
| | TEST SET | ACA | 60.70 | 74.86 | 80.57 | 63.88 | 62.66* | 70.38 | 58.09 | 62.71 | 71.94 | 58.40 | 62.84 | 71.62 |
| | | OCA | 60.91 | 74.55 | 80.91 | 63.64 | 68.18 | 71.82 | 58.18 | 67.27 | 72.73 | 58.18 | 64.54 | 72.73 |
| $\mathcal{P}_2$ | TRAINING SET | ACA | 77.63 | 76.89* | 80.69 | 76.65 | 78.00 | 84.33 | 73.97 | 74.30 | 83.70 | 72.11 | 71.19* | 83.61 |
| | | OCA | 77.61 | 73.91* | 81.09 | 76.74 | 76.30* | 84.57 | 73.91 | 70.65* | 84.13 | 72.17 | 66.96* | 83.91 |
| | TEST SET | ACA | 68.89 | 63.53* | 76.46 | 63.57 | 62.13* | 70.49 | 60.95 | 62.32 | 75.34 | 60.40 | 54.60* | 71.21 |
| | | OCA | 69.09 | 61.82* | 74.55 | 62.73 | 61.82* | 70.91 | 60.91 | 60.00* | 75.46 | 60.00 | 51.82* | 71.82 |
| $\mathcal{P}_3$ | TRAINING SET | ACA | 74.46 | 88.94 | 93.18 | 73.33 | 83.85 | 87.19 | 67.95 | 75.81 | 81.63 | 69.59 | 76.46 | 81.80 |
| | | OCA | 74.56 | 87.61 | 91.52 | 73.26 | 82.17 | 86.52 | 68.04 | 75.22 | 81.52 | 69.78 | 71.96 | 81.52 |
| | TEST SET | ACA | 62.63 | 72.71 | 83.44 | 61.80 | 68.18 | 83.95 | 58.62 | 59.49 | 75.38 | 58.57 | 64.03 | 70.75 |
| | | OCA | 61.82 | 73.64 | 83.80 | 60.91 | 70.00 | 81.82 | 58.18 | 61.82 | 76.36 | 57.27 | 63.64 | 70.91 |
| $\mathcal{P}_4$ | TRAINING SET | ACA | 77.67 | 73.84* | 87.05 | 77.51 | 74.26* | 84.40 | 73.89 | 71.35* | 82.93 | 74.32 | 72.12* | 81.25 |
| | | OCA | 77.61 | 74.13* | 87.82 | 77.61 | 75.00* | 84.54 | 73.91 | 71.30* | 82.61 | 74.35 | 72.17* | 80.65 |
| | TEST SET | ACA | 66.46 | 59.70* | 77.33 | 63.98 | 64.28 | 75.82 | 60.44 | 53.00* | 73.43 | 63.71 | 60.11* | 72.90 |
| | | OCA | 66.36 | 60.00* | 78.18 | 62.73 | 63.64 | 75.46 | 60.00 | 52.73* | 74.55 | 63.64 | 60.00* | 72.73 |
| $\mathcal{P}_5$ | TRAINING SET | ACA | 75.16 | 83.34 | 88.24 | 72.61 | 75.34 | 81.45 | 65.76 | 71.28 | 78.24 | 70.04 | 74.58 | 78.81 |
| | | OCA | 75.22 | 80.00 | 86.52 | 72.61 | 75.65 | 80.87 | 65.65 | 75.87 | 79.56 | 70.00 | 73.48 | 78.70 |
| | TEST SET | ACA | 61.76 | 74.36 | 81.12 | 66.08 | 68.50 | 77.47 | 55.55 | 55.33* | 65.44 | 57.63 | 57.22* | 64.97 |
| | | OCA | 61.82 | 71.82 | 79.09 | 65.45 | 65.45 | 74.54 | 55.45 | 60.00 | 66.36 | 58.18 | 60.91 | 67.27 |
| $\mathcal{P}_6$ | TRAINING SET | ACA | 80.60 | 84.69 | 89.39 | 73.13 | 82.78 | 87.16 | 72.57 | 78.41 | 86.09 | 74.90 | 80.86 | 86.68 |
| | | OCA | 80.65 | 85.00 | 89.78 | 73.26 | 83.04 | 87.17 | 72.39 | 78.91 | 86.09 | 74.78 | 81.96 | 86.52 |
| | TEST SET | ACA | 65.76 | 67.25 | 76.95 | 57.44 | 67.70 | 72.01 | 56.94 | 68.33 | 76.47 | 60.88 | 68.17 | 79.79 |
| | | OCA | 65.46 | 67.27 | 76.37 | 56.36 | 66.37 | 71.82 | 57.27 | 67.27 | 75.46 | 60.91 | 70.00 | 80.00 |
| $\mathcal{P}_7$ | TRAINING SET | ACA | 73.77 | 86.46 | 87.37 | 68.45 | 75.68 | 80.49 | 64.71 | 69.07 | 75.44 | 72.67 | 72.07* | 79.39 |
| | | OCA | 73.91 | 85.22 | 89.35 | 68.48 | 73.26 | 79.13 | 64.78 | 70.44 | 76.09 | 72.61 | 68.91* | 78.70 |
| | TEST SET | ACA | 62.51 | 68.95 | 77.07 | 55.65 | 64.75 | 76.66 | 58.96 | 57.56* | 67.13 | 61.10 | 52.12* | 66.90 |
| | | OCA | 61.82 | 74.54 | 79.09 | 54.55 | 63.64 | 74.55 | 58.18 | 63.64 | 67.27 | 60.91 | 54.55* | 65.45 |
| $\mathcal{P}_8$ | TRAINING SET | ACA | 78.99 | 82.55 | 91.64 | 72.01 | 78.05 | 86.49 | 72.86 | 74.80 | 80.13 | 74.20 | 74.37 | 85.28 |
| | | OCA | 78.91 | 85.65 | 90.44 | 72.17 | 73.48 | 82.61 | 72.83 | 73.70 | 78.70 | 74.13 | 73.70* | 81.31 |
| | TEST SET | ACA | 69.03 | 71.81 | 80.24 | 60.25 | 65.99 | 72.04 | 59.73 | 63.62 | 70.53 | 61.91 | 61.28* | 76.59 |
| | | OCA | 69.09 | 77.27 | 83.64 | 58.18 | 64.55 | 73.64 | 59.09 | 67.27 | 72.73 | 61.82 | 63.64 | 73.64 |

Data on average and overall accuracy are expressed in percentage.

According to the obtained results, the average and the overall accuracy rates decrease in years prior the financial distress, underlying that the model becomes less efficient with years in replicating a pre-specified classification. This trend is more evident in M.H.DIS model developed with AMADEUS classification than the one obtained with PROMETHEE method, especially for pairs $\mathcal{P}_2$ and $\mathcal{P}_3$.

Moreover, the higher performances of M.H.DIS model developed with PROMETHEE-based classification in terms of accuracy rate is generally confirmed in the same previous combinations of criteria referred to subsets $F_h^c$ with $h = 1, 3, 5, 6, 7$ and 8. Indeed, in these last combinations ACA and OCA of M.H.DIS model performed with PROMETHEE-based classification are always higher than the one achieved with AMADEUS classification, regardless the preference functions used except for the ACA of the following test set: $\mathcal{P}_1$ for year-2; $\mathcal{P}_5$ for year-3, year-4; and $\mathcal{P}_8$ for year-4.

Similarly, in combinations $\mathcal{P}_2$ and $\mathcal{P}_4$, the M.H.DIS model built on AMADEUS database, displays an ACA and OCA that are within intervals of the minimum and maximum value attained through the multi-criteria discrimination model built with PROMETHEE method also for year-2, year-3, year-4, confirming the results obtained in year-1.

Unclear case is combination $\mathcal{P}_7$, where the results of year-4 are opposite to year-1. Specifically, the discrimination model performed with AMADEUS classification achieves in this last year, an accuracy rate that is higher than the minimum accuracy value obtained with PROMETHEE based classification in both training and test set, but never higher than its maximum, giving however robustness to the PROMETHEE method in classifying companies also for previous years to financial distress.

## 8. CONCLUSIONS

In light of the recent flawed risk management actions of banks and deregulation processes introduced in the European energy industry on December 1996, the development and use of more reliable and accurate failure prediction models is becoming of major importance for energy companies, in order to prevent financial repercussions that could be catastrophic for the economy of a country.

While several statistical techniques are widely employed to deal with the issue of companies' credit risk assessment, multicriteria models are often preferred to them thanks to their high comprehensibility, easiness of application and ability to incorporate the DM's preferences.

Thus, this study employs one of the most efficient multi-criteria failure prediction models, the Multi-group Hierarchy Discrimination (M.H.DIS) technique elaborated by Zopounidis and Doumpos (2000). It has been applied on a balanced sample of 114 active and inactive European unlisted energy companies for up to four years prior the financial distress occurred. Moreover, in order to avoid the issue of small sample and to develop the model adequately, a five-fold cross validation has been performed to analyze whether the pre-specified classification of companies provided by Amadeus database is well replicated by the model.

Since the M.H.DIS method achieves a quite limited satisfactory accuracy in predicting the considered Amadeus classification in the holdout sample (68.18%), the PROMETHEE method has been performed then to provide a benchmark sorting procedure useful for comparison purposes. Thus, the six financial variables, previously selected to implement the M.H.DIS model with AMADEUS based classification, have been considered in eight combinations and employed in turn in subsets of three criteria in the building of PROMETHEE classification first and M.H.DIS model development then. The evidences provided in this paper highlight the robustness of M.H.DIS model developed with PROMETHEE based classification as consequence of the following three main results:

(1) by considering all possible combinations of more powerful financial variables in well distinguishing the two classes, the discrimination power of M.H.DIS model developed with PROMETHEE based classification in year-1 is higher than the one obtained with AMADEUS classification on six of the eight pairs $\mathcal{P}_h$ with $h = 1, 3, 5, 6, 7$ and $8$ for training and test set;

(2) by taking into account the whole set of preference functions to build a PROMETHEE based classification, it is worthy to note that PROMETHEE model represents a robust tool to sort companies into categories since the majority of preference functions classify companies into the same healthiest and riskiest class. Moreover, the results of the M.H.DIS model developed with PROMETHEE based classification show a higher performance in terms of accuracy rate than AMADEUS one, regardless of the preference function used. Indeed, in all combinations, the performance indicators relative to AMADEUS based classification are never higher than the maximum accuracy value achieved with the six preference functions used in PROMETHEE based classification;

(3) by simulating the weights of criteria in 10,000 different scenarios with the hit and run procedure, the final PROMETHEE based classification handles with the DM's uncertainty on criteria weights providing a more robust assessment of the companies' classification. This is further confirmed by the fact that cases with the highest accuracy rate ($\mathcal{P}_h$ with $h = 1, 3, 5, 6, 7$ and $8$) share common features such as: the not equally sample distribution between the two classes with a concentration in favor of class $C_1$, the attainment of the lowest performance results where the preference functions is more complex (level, linear or Gaussian criterion), the development of the M.H.DIS model on at least two financial variables with a greater weight in discriminating between class $C_1$ and $C_2$ (Table 6).

Moreover, if on the one side the efficiency of the M.H.DIS model decreases with years (year-2, year-3, year-4), on the other side, the robustness of PROMETHEE based classification against the AMADEUS one is further confirmed in the same aforementioned combinations ($\mathcal{P}_h$ with $h = 1, 3, 5, 6, 7$ and $8$) and regardless of the preference function employed, even for years before financial distress occurred. Indeed, similarly to year-1, in all combinations of year-2, year-3 and year-4, the average and the overall accuracy rate of M.H.DIS model developed with AMADEUS based classification never exceed the maximum accuracy value obtained with the six preference functions employed in PROMETHEE based classification, taking otherwise an intermediate value within the range of possible six accuracy values.

Therefore, the noteworthy results obtained in this study show that PROMETHEE based classification, used jointly with M.H.DIS model, enhances the performances of the discrimination model specifically for credit risk assessment of energy companies. More generally, this approach is recommended in two cases:

- whenever the M.H.DIS model developed with a pre-specified classification give results not fully satisfactory in terms of overall accuracy;
- whenever the sample under consideration is composed by alternatives for which the credit rating are not provided by credit rating agencies (CRAs) as in the case of unlisted or small and medium-sized enterprises (SMEs), even if a support to the credit risk assessment process is relevant also in this case.

Future research could also focus on extending the proposed methodology with respect:

- <u>to variables</u>: in this study, we employed only financial variables to evaluate the creditworthiness of energy companies. Future researches could be devoted to investigate whether soft variables, such as management, market and macro-economic variables, could affect the creditworthiness of energy companies or could improve the predictive accuracy of the distress model.

Moreover, it might be interesting to consider the multidimensional nature of the energy companies' assessment that requires the definition of a hierarchical structure of criteria including elements such as the environmental, the technical and the market criteria, to observe whether the accuracy of the two combined models performed on this evaluation could increase against externally assigned ratings;

- to the dataset: in this study, unlisted European energy companies composed the sample and the lack of synthetic rating judgement provided by CRAs has been highlighted for this sample. Therefore, a possible future direction could consider a set of alternatives composed by listed energy companies in order to compare the results obtained through our proposed methodology, i.e. the combination of M.H.DIS and PROMETHEE II model, and the M.H.DIS developed with the pre-defined classification issued by credit rating agencies (CRAs). In this way, it could be possible to observe which of two classifications is better replicated by the discrimination model.

**Acknowledgments**

The authors would like to thank Prof. Michalis Doumpos to provide us the MATLAB code for the M.H.DIS method.

For the entire research process of this study, the authors have benefited the PRD fund of the University of Catania "Indebtedness, bank credit and economic activities". The first author has also benefited the "FFABR" fund of the Ministry of Education, University and Research of the Italian government.

# Appendix A

Table 1A. Energy companies in the final balanced sample after the stratified resampling method distributed per country. Authors' elaboration

| COUNTRY | ELECTRIC COMPANIES | | | |
|---|---|---|---|---|
| | ACTIVE | Relative frequency | INACTIVE | Relative frequency |
| GERMANY | STADTWERKE BONN GMBH (SWB)<br>STADTWERKE WEIßENBURG GMBH<br>STADTWERKE PRENZLAU GMBH<br>STADTWERKE SCHWEINFURT GMBH<br>STADTWERKE GREIFSWALD GESELLSCHAFT MIT BESCHRÄNKTER HAFTUNG<br>ENERGIEVERSORGUNG SEHNDE GMBH<br>ENERGIEEINKAUFS- UND -HANDELSGESELLSCHAFT MECKLENBURG-VORPOMMERN MBH<br>STADTWERKE EBERBACH<br>STADTWERKE HUSUM GMBH<br>STADTWERKE WERL GMBH<br>STADTWERKE ERDING GMBH | 19.30% | EEV BIOENERGIE GMBH & CO. KG<br>OVAG ENERGIE AG<br>MT-BIOMETHAN GMBH | 5.26% |
| SPAIN | SUN EUROPEAN INVESTMENTS EOLICO OLIVILLO SA.<br>MOLINOS DEL EBRO SA<br>CONTOURGLOBAL LA RIOJA SL<br>EVOLUCION 2000 SOCIEDAD LIMITADA.<br>M TORRES DESARROLLOS ENERGETICOS SL<br>SOLYNOVA VALVERDON SL<br>BIO OILS ENERGY SA<br>TECNOHUERTAS SA<br>GRANSOLAR DESARROLLO Y CONSTRUCCION SL. | 15.79% | SIBERIA SOLAR SL<br>SERRA DO MONCOSO-CAMBAS SL<br>X-ELIO REAL ESTATE ENERGY SL.<br>ALTEN POZOHONDO SOCIEDAD LIMITADA<br>PARQUE SOLAR LA ROBLA SL<br>ALTEN ALANGE SL<br>AUDAX ENERGIA SA<br>PLANSOFOL SL | 8.77% |
| ITALY | C.V.A. VENTO S.R.L.<br>SOCIETA' ELETTRICA IN MORBEGNO SOCIETA' COOPERATIVA PER AZIONI<br>EOLICA SANTOMENNA S.R.L.<br>ERMES GAS & POWER SOCIETA' A RESPONSABILITA' LIMITATA<br>SOCIETA' ENERGIE RINNOVABILI 1 SOCIETA' PER AZIONI<br>ENOMONDO S.R.L.<br>AGSM ENERGIA S.P.A.<br>ORSA MAGGIORE PV S.R.L.<br>ALPERIA VIPOWER SPA<br>ENERGIA UNO S.R.L.<br>TG MASSERIA GIORGINI S.R.L.<br>IMPIANTO ALPHA S.R.L.<br>OTTANA SOLAR POWER S.P.A. | 22.80% | EVIVA S.P.A. IN LIQUIDAZIONE<br>ELECTRA ITALIA S.P.A.<br>ENERGHE S.P.A.<br>TRADECOM S.P.A<br>E.S.TR.A. ELETTRICITA' S.P.A.<br>AEVV ENERGIE S.R.L.<br>ENERGIA E TERRITORIO – SRL<br>AZIENDA ENERGETICA VALTELLINA VALCHIAVENNA S.P.A.<br>AP ENERGIA S.R.L. - IN LIQUIDAZIONE<br>ESPERIA SOCIETA' PER AZIONI IN LIQUDAZIONE<br>HOLDING FORTORE ENERGIA S.R.L.<br>LINEA RETI E IMPIANTI S.R.L.<br>UNIPOWER ITALIA S.R.L.<br>GENERAL POWER S.R.L. IN LIQUIDAZIONE<br>HELIOS ITA 3 S.R.L.<br>SOLAR ENERGY ITALIA 7 SRL<br>EMMECIDUE S.R.L. IN LIQUIDAZIONE<br>VENUSIA SRL<br>PARCO EOLICO GIRIFALCO S.R.L. | 52.63% |
| | | | VARSI FOTOVOLTAICO SRL
GREENSOURCE S.P.A.
S5 SRL
EN & EN - ENERGIE PER ENERGIA S.R.L.
EF AUGUSTA S.R.L.
VILLA CASTELLI WIND S.R.L.
IDREG-PIEMONTE - S.P.A.
ITALBREVETTI SOCIETA' A RESPONSABILITA' LIMITATA
STS SOCIETA' TERMOELETTRICA SEDRINA S.R.L
SUNSHIRE S.R.L.
FILOVERDE S.P.A. | |
| FRANCE | CENTRALE EOLIENNE DE PRODUCTION D'ENERGIE DE HAUT CHEMIN
EWZ PARC EOLIEN EPINETTE
ELICIO VENT D'OUEST | 5.26% | ENGIE NUCLEAR DEVELOPMENT
FORCES HYDRAULIQUES DE MEUSE
LA COMPAGNIE DU VENT
ALBIOMA CARAIBES | 7.01% |
| SWEDEN | KRISTINEHAMNS ELNÄT AB
SKÅNSKA ENERGI NÄT AKTIEBOLAG
HÄRRYDA ENERGI AKTIEBOLAG
AB BORLÄNGE ENERGI ELNÄT | 7.01% | | 0% |
| FINLAND | VOIMAPATO OY
PARIKKALAN VALO OY
LAPPEENRANNAN ENERGIAVERKOT OY | 5.26% | | 0% |
| GREECE | GREEK ENVIRONMENTAL & ENERGY NETWORK A.E.
ΗΡΩΝ ΘΕΡΜΟΗΛΕΚΤΡΙΚΗ Α.Ε. | 3.50% | | 0% |
| DANMARK | VESTJYSKE NET 60 KV A/S
GRINDSTED EL- OG VARMEVÆRK A.M.B.A | 3.50% | | 0% |
| ROMANIA | OET ROMANIA LTD BULGARIA SUCURSALA BUCURESTI

SOCIETATEA DE DISTRIBUȚIE A ENERGIEI ELECTRICE TRANSILVANIA SUD. | 3.50% | SOCIETATEA COMERCIALA DE PRODUCERE A ENERGIEI ELECTRICE SI TERMICE "TERMOELECTRICA" | 1.75% |
| PORTUGAL | TEJO ENERGIA - PRODUÇÃO E DISTRIBUIÇÃO DE ENERGIA ELÉCTRICA, S.A.
BIOELÉCTRICA DA FOZ, S.A. | 3.50% | | 0% |
| BULGARIA | ТОПЛОФИКАЦИЯ РУСЕ ЕАД
ЕЛЕКТРОЕНЕРГИЕН СИСТЕМЕН ОПЕРАТОР ЕАД | 3.50% | ТОПЛОФИКАЦИЯ ПЕТРИЧ ЕАД
ЕНЕРГИЙНА ФИНАНСОВА ГРУПА АД | 3.50% |
| BELGIUM | ESSENT BELGIUM | 1.75% | INTERCOMMUNALE MAATSCHAPPIJ VOOR ENERGIEVOORZIENING ANTWERPEN | 1.75% |
| SLOVENIA | ELEKTRO MARIBOR, PODJETJE ZA DISTRIBUCIJO ELEKTRIČNE ENERGIJE, D.D. | 1.75% | | 0% |
| LATVIA | AUGSTSPRIEGUMA TĪKLS AS | 1.75% | | 0% |
| POLAND | EOLOS POLSKA SP. Z O.O. | 1.75% | PARK WIATROWY TYCHOWO SP. Z O.O.
PARK WIATROWY NOWY STAW SP. Z O.O. | 3.50% |
| HUNGARY | | 0% | VEOLIA SZOLGÁLTATÓ KÖZPONT MAGYARORSZÁG KFT
MISTRAL ENERGETIKA VILLAMOSENERGIA-TERMELŐ KFT
KAPTÁR SZÉLERŐMŰ KERESKEDELMI ÉS SZOLGÁLTATÓ KFT
MVM ÉSZAK-BUDAI KOGENERÁCIÓS FŰTŐERŐMŰ KFT | 7.01% |
| CZECHIA | | 0% | MORAVIA GREEN POWER S.R.O. | 1.75% |
| SLOVAKIA | | 0% | LUMIUS SLOVAKIA, S.R.O. V LIKVIDÁCII | 1.75% |

Table 2A. Stages performed to select independent variables of the sample introduced in the failure prediction model. Authors' elaboration.

| INDEPENDENT VARIABLES (FRs) | | Stage 1 | | | Stage 2 | | Stage 3 |
|---|---|---|---|---|---|---|---|
| PROFITABILITY | Year | *IV* Value | Predictive power | | t-test p value | Predictive power | Correlation analysis |
| EBIT_TA | -1 | 0.7226 | SUSPICIOUS | S | 0.4934 | NS | NS |
| | -2 | 0.4472 | STRONG | S | 0.0123 | S | NS |
| | -3 | 0.1583 | MEDIUM | S | 0.1536 | NS | NS |
| | -4 | 0.4165 | STRONG | S | 0.4381 | NS | NS |
| LTDR | -1 | 0.4621 | STRONG | S | 0.1904 | NS | NS |
| | -2 | 0.2118 | MEDIUM | S | 0.4909 | NS | NS |
| | -3 | 0.1907 | MEDIUM | S | 0.6455 | NS | NS |
| | -4 | 0.0825 | WEAK | NS | 0.8063 | NS | NS |
| OP_MARG | -1 | 0.8934 | SUSPICIOUS | S | 0.2695 | NS | NS |
| | -2 | 0.2635 | STRONG | S | 0.0219 | S | NS |
| | -3 | 0.097 | WEAK | NS | 0.4368 | NS | NS |
| | -4 | 0.3002 | STRONG | S | 0.1694 | NS | NS |
| PROF_MARG | -1 | 10.343 | SUSPICIOUS | S | 0.0223 | S | NS |
| | -2 | 0.4354 | STRONG | S | 0.0284 | S | NS |
| | -3 | 0.2902 | STRONG | S | 0.4255 | NS | NS |
| | -4 | 0.324 | STRONG | S | 0.1405 | NS | NS |
| ROE | -1 | 0.6102 | SUSPICIOUS | S | 0.0223 | S | NS |
| | -2 | 0.2479 | MEDIUM | S | 0.7253 | NS | NS |
| | -3 | 0.2014 | MEDIUM | S | 0.683 | NS | NS |
| | -4 | 0.0961 | WEAK | NS | 0.2315 | NS | NS |
| ROA | -1 | 10.723 | SUSPICIOUS | S | 0.0018 | S | S |
| | -2 | 0.4323 | STRONG | S | 0.0088 | S | S |
| | -3 | 0.2449 | MEDIUM | S | 0.0717 | S | S |
| ROCE | -4 | 0.4779 | STRONG | S | 0.05 | S | S |
| | -1 | 0.8736 | SUSPICIOUS | S | 0.1202 | NS | NS |
| | -2 | 0.4537 | STRONG | S | 0.1502 | NS | NS |
| | -3 | 0.128 | MEDIUM | S | 0.9242 | NS | NS |
| | -4 | 0.2545 | STRONG | S | 0.9431 | NS | NS |
| EBIT_EQ | -1 | 0.4227 | STRONG | S | 0.6235 | NS | NS |
| | -2 | 0.3912 | STRONG | S | 0.3479 | NS | NS |
| | -3 | 0.3932 | STRONG | S | 0.2048 | NS | NS |
| | -4 | 0.2014 | MEDIUM | S | 0.2434 | NS | NS |
| EBITDA_TA | -1 | 0.4583 | STRONG | S | 0.448 | NS | S |
| | -2 | 0.7929 | SUSPICIOUS | S | 0.0053 | S | S |
| | -3 | 0.5576 | SUSPICIOUS | S | 0.0275 | S | S |
| | -4 | 0.5049 | SUSPICIOUS | S | 0.049 | S | S |
| CF_TA | -1 | 0.769 | SUSPICIOUS | S | 0.0012 | S | NS |
| | -2 | 0.6866 | SUSPICIOUS | S | 0.007 | S | NS |
| | -3 | 0.5734 | SUSPICIOUS | S | 0.659 | NS | NS |
| | -4 | 0.7187 | SUSPICIOUS | S | 0.0008 | S | NS |
| CF_EQ | -1 | 0.4227 | STRONG | S | 0.0877 | S | NS |
| | -2 | 0.3981 | STRONG | S | 0.3296 | NS | NS |
| | -3 | 0.2279 | MEDIUM | S | 0.271 | NS | NS |
| | -4 | 0.3912 | STRONG | S | 0.2598 | NS | NS |
| **FINANCIAL STRUCTURE** | | | | | | | |
| EQ_RATIO | -1 | 0.4235 | STRONG | S | 0.0023 | S | S |
| | -2 | 0.5111 | SUSPICIOUS | S | 0.0131 | S | S |
| | -3 | 0.459 | STRONG | S | 0.0229 | S | S |
| | -4 | 0.3624 | STRONG | S | 0.0747 | S | S |
| FAT | -1 | 0.1932 | MEDIUM | S | 0.23 | NS | NS |
| | -2 | 0.1875 | MEDIUM | S | 0.5129 | NS | NS |
| | -3 | 0.1586 | MEDIUM | S | 0.7108 | NS | NS |
| | -4 | 0.2236 | MEDIUM | S | 0.8005 | NS | NS |
| TD_TA | -1 | 0.4185 | STRONG | S | 0.0007 | S | S |
| | -2 | 0.5272 | SUSPICIOUS | S | 0.0022 | S | S |
| | -3 | 0.0035 | USELESS | NS | 0.0071 | S | S |
| | -4 | 0.4218 | STRONG | S | 0.0051 | S | S |
| LTD_EQ | -1 | 0.6287 | SUSPICIOUS | S | 0.0488 | S | NS |
| | -2 | 0.2837 | STRONG | S | 0.227 | NS | NS |
| | -3 | 0.3403 | STRONG | S | 0.2849 | NS | NS |
| | -4 | 0.0828 | WEAK | NS | 0.487 | NS | NS |
| NOWC | -1 | 0.9949 | SUSPICIOUS | S | 0.0006 | S | NS |
| | -2 | 0.244 | MEDIUM | S | 0.0267 | S | NS |
| | -3 | 0.1328 | MEDIUM | S | 0.2129 | NS | NS |
| | -4 | 0.1017 | MEDIUM | S | 0.1974 | NS | NS |
| TD_EQ | -1 | 0.2479 | MEDIUM | S | NS | NS | NS |
| | -2 | 0 | USELESS | NS | NS | NS | NS |
| | -3 | 0 | USELESS | NS | NS | NS | NS |
| | -4 | 0.0946 | WEAK | NS | NS | NS | NS |
| **LIQUIDITY** | | | | | | | |
| CA_TA | -1 | 0.1946 | MEDIUM | S | 0.0798 | S | S |
| | -2 | 0.2994 | STRONG | S | 0.0467 | S | S |
| | -3 | 0.2646 | STRONG | S | 0.0494 | S | S |
| | -4 | 0.024 | USELESS | NS | 0.1674 | NS | S |
| CR | -1 | 0.5679 | SUSPICIOUS | S | 0.4555 | NS | NS |
| | -2 | 0.1948 | MEDIUM | S | 0.3666 | NS | NS |
| | -3 | 0.0887 | WEAK | NS | 0.1555 | NS | NS |
| | -4 | 0.2882 | STRONG | S | 0.4258 | NS | NS |
| DR | -1 | 0.4235 | STRONG | S | 0.0023 | S | NS |
| | -2 | 0.5111 | SUSPICIOUS | S | 0.0131 | S | NS |
| | -3 | 0.459 | STRONG | S | 0.0229 | S | NS |
| | -4 | 0.3624 | STRONG | S | 0.0747 | S | NS |
| WC_TA | -1 | 0.5326 | SUSPICIOUS | S | 0.0175 | S | NS |
| | -2 | 0.5181 | SUSPICIOUS | S | 0.0526 | S | NS |
| | -3 | 0.128 | MEDIUM | S | 0.8698 | NS | NS |
| | -4 | 0.3182 | STRONG | S | 0.3793 | NS | NS |
| CASH_CL | -1 | 0.5987 | SUSPICIOUS | S | 0.009 | S | NS |
| | -2 | 0.3538 | STRONG | S | 0.0206 | S | NS |
| | -3 | 0.2226 | MEDIUM | S | 0.2166 | NS | NS |
| | -4 | 0.3615 | STRONG | S | 0.5139 | NS | NS |
| CASH_TA | -1 | 0.3164 | STRONG | S | 0.1324 | NS | NS |
| | -2 | 0.1141 | MEDIUM | S | 0.1524 | NS | NS |
| | -3 | 0.1137 | MEDIUM | S | 0.1594 | NS | NS |
| | -4 | 0.123 | MEDIUM | S | 0.8282 | NS | NS |
| CL_TA | -1 | 0.0035 | USELESS | NS | NS | NS | NS |
| | -2 | 0.0035 | USELESS | NS | NS | NS | NS |
| | -3 | 0.0946 | WEAK | NS | NS | NS | NS |
| | -4 | 0.0065 | USELESS | NS | NS | NS | NS |
| CASH_CA | -1 | 0.3685 | STRONG | S | 0.0073 | S | NS |
| | -2 | 0.2922 | STRONG | S | 0.0478 | S | NS |
| | -3 | 0.1981 | MEDIUM | S | 0.1055 | NS | NS |
| | -4 | 0.1999 | MEDIUM | S | 0.4105 | NS | NS |
| CF_CL | -1 | 10.513 | SUSPICIOUS | S | 0.1834 | NS | NS |
| | -2 | 11.251 | SUSPICIOUS | S | 0.2161 | NS | NS |
| | -3 | 0.6668 | SUSPICIOUS | S | 0.4845 | NS | NS |
|  |  | -4 | 0.6798 | SUSPICIOUS | S | 0.1408 | NS | NS |
| **SOLVENCY** |  |  |  |  |  |  |  |
| FE_EBITDA | -1 | 0.6828 | SUSPICIOUS | S | 0.4209 | NS | NS |
|  | -2 | 0.5668 | SUSPICIOUS | S | 0.796 | NS | NS |
|  | -3 | 0.0946 | WEAK | NS | 0.0106 | S | NS |
|  | -4 | 0.2036 | MEDIUM | S | 0.1563 | NS | NS |
| FE_NI | -1 | 0.9736 | SUSPICIOUS | S | 0.4022 | NS | NS |
|  | -2 | 0.2189 | MEDIUM | S | 0.8175 | NS | NS |
|  | -3 | 0.038 | USELESS | NS | 0.1887 | NS | NS |
|  | -4 | 0.5069 | SUSPICIOUS | S | 0.0711 | NS | NS |
| FE_TA | -1 | 0.0957 | WEAK | NS | NS | NS | NS |
|  | -2 | 0.0329 | USELESS | NS | NS | NS | NS |
|  | -3 | 0.0329 | USELESS | NS | NS | NS | NS |
|  | -4 | 0.0946 | WEAK | NS | NS | NS | NS |
| **TURNOVER** |  |  |  |  |  |  |  |
| CA_TS | -1 | 0.0325 | USELESS | NS | 0.0271 | S | S |
|  | -2 | 0.4823 | STRONG | S | 0.0146 | S | S |
|  | -3 | 0.4127 | STRONG | S | 0.0444 | S | S |
|  | -4 | 0.1295 | MEDIUM | S | 0.1626 | NS | S |
| CL_TS | -1 | 0.0035 | USELESS | NS | 0.0748 | S | NS |
|  | -2 | 0.4745 | STRONG | S | 0.1159 | NS | NS |
|  | -3 | 0.3306 | STRONG | S | 0.0663 | S | NS |
|  | -4 | 0.3152 | STRONG | S | 0.2084 | NS | NS |
| WC_TS | -1 | 0.5606 | SUSPICIOUS | S | 0.6022 | NS | NS |
|  | -2 | 0.1321 | MEDIUM | S | 0.7 | NS | NS |
|  | -3 | 0.1292 | MEDIUM | S | 0.2822 | NS | NS |
|  | -4 | 0.2479 | MEDIUM | S | 0.3865 | NS | NS |
| **ACTIVITY** |  |  |  |  |  |  |  |
| CF_NS | -1 | 0.6786 | SUSPICIOUS | S | 0.3012 | NS | NS |
|  | -2 | 0.7162 | SUSPICIOUS | S | 0.03 | S | NS |
|  | -3 | 0.3497 | STRONG | S | 0.455 | NS | NS |
|  | -4 | 0.3488 | STRONG | S | 0.5297 | NS | NS |
| GROW_TA | -1 | 0.4537 | STRONG | S | 0.51 | NS | NS |
|  | -2 | 0.5116 | SUSPICIOUS | S | 0.9802 | NS | NS |
|  | -3 | 0.0387 | USELESS | NS | 0.3038 | NS | NS |
|  | -4 | 0.0961 | WEAK | NS | 0.5249 | NS | NS |
| EBITDA_TS | -1 | 0.2189 | MEDIUM | S | 0.1205 | NS | NS |
|  | -2 | 0.5168 | SUSPICIOUS | S | 0.0653 | S | NS |
|  | -3 | 0.2806 | STRONG | S | 0.5821 | NS | NS |
|  | -4 | 0.2871 | STRONG | S | 0.2843 | NS | NS |
| **SIZE** |  |  |  |  |  |  |  |
| TA | -1 | 0.1802 | MEDIUM | S | 0.2417 | NS | NS |
|  | -2 | 0.1318 | MEDIUM | S | 0.2944 | NS | NS |
|  | -3 | 0.1069 | MEDIUM | S | 0.3636 | NS | NS |
|  | -4 | 0.1009 | MEDIUM | S | 0.323 | NS | NS |
| SALES | -1 | 0.2152 | MEDIUM | S | 0.5979 | NS | NS |
|  | -2 | 0.2719 | STRONG | S | 0.9897 | NS | NS |
|  | -3 | 0.2236 | MEDIUM | S | 0.7313 | NS | NS |
|  | -4 | 0.2776 | STRONG | S | 0.8508 | NS | NS |